# An Exact Solution of the 3-D Navier-Stokes Equation

A. Muriel*

Department of Electrical Engineering

Columbia University

and

Department of Philosophy

Harvard University

Abstract

We continue our work reported earlier (A. Muriel and M. Dresden, Physica D 101, 299, 1997) to calculate the time evolution of the one-particle distribution function. An improved operator formalism, heretofore unexplored, is used for uniform initial data. We then choose a Gaussian pair potential between particles. With these two conditions, the velocity fields, energy and pressure are calculated exactly. All stipulations of the Clay Mathematics Institute for proposed solutions of the 3-D Navier-Stokes Equation are satisfied by our time evolution equation solution. We then substitute the results for the velocity fields into the 3-d Navier-Stokes Equation and calculate the pressure. The results from our time evolution equation and the prescribed pressure from the Navier-Stokes Equation constitute an exact solution to the Navier-Stokes Equation. No turbulence is obtained from the solution. A philosophical discussion of the results, and their meaning for the problem of turbulence concludes this study.

PACS nos. 51.10.+y , 05.20.Dd , 05.60.-k , 05.20.Jj

Keywords: Time evolution equations, Navier-Stokes equation, exact solutions

*E-mail address: amadormuriel@fas.harvard.ed

1. Motivation

There has not been any published solution of the 3-D Navier-Stokes Equation (NSE). The purpose of this Report is to fully document a procedure for arriving at an exact solution of this well-known problem. The Report is written to satisfy the requirements of the Clay Institute of Mathematics and makes the solution available to mathematicians, physicists, and engineers who may wish to verify the claim. As my solution comes from the world of non-equilibrium statistical mechanics, the report includes some introduction that will be needed to follow the proposed solution.

2. Introduction

In a paper on an integral formulation of hydrodynamics [1], to which we refer the reader for symbols and conventions, the following exact formal result for the time evolution of the single particle distribution was derived:

$$f(r,p,t) = f(r - pt/m, p, 0) + n_o \int_0^t ds_1 \int dp' e^{-s_1 L_o} \int dr' \left( \frac{\partial V(r-r')}{\partial r} \right) \frac{\partial}{\partial p} f_2^1(r,r',p,0)$$

$$+ n_o \int_0^t ds_1 \int_0^{s_1} ds_2 \int dp' e^{-s_1 L_o} \int dr' \left( \frac{\partial V(r-r')}{\partial r} \right) \frac{\partial}{\partial p} \left( \frac{p}{m} \frac{\partial}{\partial r} \right) f_2^1(r,r',p,0)$$

$$- n_o \int_0^t ds_1 \int_0^{s_1} ds_2 \int dp' e^{-s_1 L_o} \int dr' \left( \frac{\partial V(r-r')}{\partial r} \right) \frac{\partial}{\partial p} \left( \frac{p}{m} \frac{\partial}{\partial r} \right) f_2^2(r,r',p,p'0)$$

$$+ \frac{n_o^2}{2} \int_0^t ds_1 \int_0^{s_1} ds_2 e^{-s_2 L_o} \int dr' \int dr'' \frac{\partial V(r-r')}{\partial r} \frac{\partial}{\partial p} \left( \frac{\partial V(r-r'')}{\partial r} \frac{\partial}{\partial p} f_3^1(r,r',r'',p,o) \right)$$

$$+ \frac{n_o^2}{2} \int_0^t ds_1 \int_0^{s_1} ds_2 e^{-s_2 L_o} \int dr' \frac{\partial V(r-r')}{\partial r} \frac{\partial}{\partial p} \left( \frac{\partial V(r-r')}{\partial r} \frac{\partial}{\partial p} f_2^1(r,r',p,0) \right) + \sum_{n=3}^{\infty} O\left( \frac{\partial^n}{\partial p^n} \right)$$

(1)

$r, p$ refer to the Cartesian spatial coordinate and momentum of one particle. Cartesian dot products are implied. In Eq. (1) the superscripts refer to the number of momentum coordinates and the subscripts to the number of spatial coordinates. The exact solution includes an infinite series in the operator $\frac{\partial^n}{\partial p^n}$ for all $n = 3...\infty$, but as we will only calculate the average of the momentum and the energy, the formal solution will rigorously truncate to only a few terms by partial integration.

We will start with initial data which is uniform, the solution simplifies even further. Note that the integrals of force over all space is zero and what remains of Eq. (1) is

$$f(r,p,t) = f(r - pt/m, p, 0)$$
$$+ \frac{n_o^2}{2} \int_0^t ds_1 \int_0^{s_1} ds_2 e^{-s_2 L_o} \int dr' \left(\frac{\partial V(r-r')}{\partial r}\right)^2 \frac{\partial^2}{\partial p^2} f_2^1(r,r',p,0)$$

(2)

The operator $L_o$ is given by

$$L_o = -\left(\frac{p_x}{m}\frac{\partial}{\partial x} + \frac{p_y}{m}\frac{\partial}{\partial y} + \frac{p_z}{m}\frac{\partial}{\partial z}\right) \tag{3}$$

Notice that the well-known BBGKY hierarchy [2] of the 1-particle distribution function coupled to the 2-particle distribution function, and the 2-particle distribution function itself coupled to the 3-particle distribution function, etc., has been replaced by initial data correlations, but only up to 3-particles.

We will evaluate (2) step by step, from the right.

3. **Uniform initial data and a one dimensional jet.**

We start with a uniform system and the following initial momentum distribution

$$\varphi(p) = \delta(p_x - p_o)\delta(p_y)\delta(p_z) \tag{4}$$

a one-dimensional jet satisfying an initial condition stipulated by the Clay Institute. We put $n_0 = 1$

Next, we choose the pair-potential

$$V(r-r') = g\exp\{-\alpha[(x-x')^2 + (y-y')^2 + (z-z')^2]\} \tag{5}$$

and integrate it over all $x', y', z'$ over a cube of dimension $L$

to give the integral over all space resulting in the display expression $h(x, y, z)$

which we redefine in Maple symbolic code as Int3DVsq , the integral over all volume of the force squared.

$$
\begin{aligned}
Int3DVsq := & \frac{1}{32} \frac{1}{\sqrt{a}} \Big( g^2 \Big( 4 \pi x \, \mathrm{erf}(\sqrt{a}\, y\, \sqrt{2}) \sqrt{a} \, e^{2aL^2} \\
& + 4 \pi x \, \mathrm{erf}(\sqrt{2}\, \sqrt{a}\, L - \sqrt{a}\, y\, \sqrt{2}) \sqrt{a} \, e^{2aL^2} \\
& - \pi^{3/2} \sqrt{2}\, \mathrm{erf}(\sqrt{a}\, x\, \sqrt{2})\, \mathrm{erf}(\sqrt{a}\, y\, \sqrt{2})\, e^{2ax^2 + 2aL^2} \\
& - \pi^{3/2} \sqrt{2}\, \mathrm{erf}(\sqrt{a}\, x\, \sqrt{2})\, \mathrm{erf}(\sqrt{2}\, \sqrt{a}\, L \\
& - \sqrt{a}\, y\, \sqrt{2})\, e^{2ax^2 + 2aL^2} \\
& + 4 \pi L \, e^{4aLx}\, \mathrm{erf}(\sqrt{a}\, y\, \sqrt{2}) \sqrt{a} \\
& + 4 \pi L \, e^{4aLx}\, \mathrm{erf}(\sqrt{2}\, \sqrt{a}\, L - \sqrt{a}\, y\, \sqrt{2}) \sqrt{a} \\
& - 4 \pi x \, e^{4aLx}\, \mathrm{erf}(\sqrt{a}\, y\, \sqrt{2}) \sqrt{a} \\
& - 4 \pi x \, e^{4aLx}\, \mathrm{erf}(\sqrt{2}\, \sqrt{a}\, L - \sqrt{a}\, y\, \sqrt{2}) \sqrt{a} \\
& + \pi^{3/2} \sqrt{2}\, \mathrm{erf}(-\sqrt{2}\, \sqrt{a}\, L \\
& + \sqrt{a}\, x\, \sqrt{2})\, \mathrm{erf}(\sqrt{a}\, y\, \sqrt{2})\, e^{2ax^2 + 2aL^2} + \pi^{3/2} \sqrt{2}\, \mathrm{erf}( \\
& -\sqrt{2}\, \sqrt{a}\, L + \sqrt{a}\, x\, \sqrt{2})\, \mathrm{erf}(\sqrt{2}\, \sqrt{a}\, L \\
& - \sqrt{a}\, y\, \sqrt{2})\, e^{2ax^2 + 2aL^2} \Big) \Big( -\mathrm{erf}(\sqrt{a}\, z\, \sqrt{2}) + \mathrm{erf}( \\
& -\sqrt{2}\, \sqrt{a}\, L + \sqrt{a}\, z\, \sqrt{2}) \Big) e^{-2ax^2 - 2aL^2} \Big)
\end{aligned}
$$

(6)

In this Report, we will be using the symbolic programming Language Maple 13, and we will shift from display equations in the tradition of physics papers, and Maple code. The symbolic code that produces (7) is given by

> gc(); restart;

> a := 'a'; g := 'g'; L := 'L';

> V := proc (x2, x) g*exp(-a*((x2-x)^2+(y2-y)^2+(z2-z)^2)) end proc;

> DVdx2 := diff(V(x2, x), x2);

> DVsq := DVdx2^2;

> IntDVsq := Int(DVsq, x2 = 0 .. L);

> IntDVsqx := int(DVsq, x2 = 0 .. L);

> IntDvsqx := simplify(IntDVsqx);

> Int2DVsqxy := int(IntDVsqx, y2 = 0 .. L);

> Int2DVsqxy := simplify(Int2DVsqxy);

> Int3DVsq := int(Int2DVsqxy, z2 = 0 .. L);

Next we use the identity

$$\int dp\, G(p) \frac{\partial^2 \delta(p - p_o)}{\partial p^2} = \frac{\partial^2 G(p)}{\partial p^2} \bigg]_{p=p_o} \tag{7}$$

where $G(p)$ is an operator which we will define below. This new operator identity is a simple but significant advance over our 1997 paper. It seems to be the first use of the Dirac-delta function with operators of the kind we propose.

### 4. Momentum time evolution

For the x-momentum we have to calculate the expression

$$\int dp_x p_x e^{-\left(\frac{p_x s_x}{m}\frac{\partial}{\partial x} + \frac{p_y s_x}{m}\frac{\partial}{\partial y} + \frac{p_z s_x}{m}\frac{\partial}{\partial z}\right)} h(x,y,z) \frac{\partial^2}{\partial p_x^2} \delta(p_x - p_o) \tag{8}$$

Using the operator identity in Eq. (7), we define the x-momentum propagator

$$Dpx2momentumpropagator :=$$
$$-\frac{2\, s2\, Dx\, e^{-\frac{px\, s2\, Dx}{m} - \frac{py\, s2\, Dy}{m} - \frac{pz\, s2\, Dz}{m}}}{m}$$
$$+ \frac{px\, s2^2\, Dx^2\, e^{-\frac{px\, s2\, Dx}{m} - \frac{py\, s2\, Dy}{m} - \frac{pz\, s2\, Dz}{m}}}{m^2} \tag{9}$$

where $D_x$ is differentiation with respect to $x$. The derivatives $Dx$ and $Dx^2$ act on shifted $h(x, y, z)$. Notice that we again exchange our display equations with Maple 13 code.

The shift operator acts on $h(x, y, z)$ to give

> *Int3DVsqshifted* ;

$$\frac{1}{32}\frac{1}{\sqrt{a}}\Bigg(g^2\Bigg(4\pi\left(x-\frac{px\,s2}{m}\right)\mathrm{erf}\left(\sqrt{a}\left(y\right.\right.$$

$$\left.\left.-\frac{py\,s2}{m}\right)\sqrt{2}\right)\sqrt{a}\,e^{2aL^2}+4\pi\left(x\right.$$

$$\left.-\frac{px\,s2}{m}\right)\mathrm{erf}\left(\sqrt{2}\sqrt{a}\,L-\sqrt{a}\left(y\right.\right.$$

$$\left.\left.-\frac{py\,s2}{m}\right)\sqrt{2}\right)\sqrt{a}\,e^{2aL^2}-\pi^{3/2}\sqrt{2}\,\mathrm{erf}\left(\sqrt{a}\left(x\right.\right.$$

$$\left.\left.-\frac{px\,s2}{m}\right)\sqrt{2}\right)\mathrm{erf}\left(\sqrt{a}\left(y\right.\right.$$

$$\left.\left.-\frac{py\,s2}{m}\right)\sqrt{2}\right)e^{2a\left(x-\frac{px\,s2}{m}\right)^2+2aL^2}$$

$$-\pi^{3/2}\sqrt{2}\,\mathrm{erf}\left(\sqrt{a}\left(x-\frac{px\,s2}{m}\right)\sqrt{2}\right)\mathrm{erf}\left(\sqrt{2}\sqrt{a}\,L\right.$$

$$\left.-\sqrt{a}\left(y-\frac{py\,s2}{m}\right)\sqrt{2}\right)e^{2a\left(x-\frac{px\,s2}{m}\right)^2+2aL^2}$$

$$+4\pi L\,e^{4aL\left(x-\frac{px\,s2}{m}\right)}\mathrm{erf}\left(\sqrt{a}\left(y-\frac{py\,s2}{m}\right)\sqrt{2}\right)\sqrt{a}$$

$$+4\pi L\,e^{4aL\left(x-\frac{px\,s2}{m}\right)}\mathrm{erf}\left(\sqrt{2}\sqrt{a}\,L-\sqrt{a}\left(y\right.\right.$$

$$\left.\left.-\frac{py\,s2}{m}\right)\sqrt{2}\right)\sqrt{a}-4\pi\left(x\right.$$

$$\left.-\frac{px\,s2}{m}\right)e^{4aL\left(x-\frac{px\,s2}{m}\right)}\mathrm{erf}\left(\sqrt{a}\left(y-\frac{py\,s2}{m}\right)\sqrt{2}\right)\sqrt{a}$$

$$-4\pi\left(x-\frac{px\,s2}{m}\right)e^{4aL\left(x-\frac{px\,s2}{m}\right)}\mathrm{erf}\left(\sqrt{2}\sqrt{a}\,L\right.$$

$$\left.-\sqrt{a}\left(y-\frac{py\,s2}{m}\right)\sqrt{2}\right)\sqrt{a}+\pi^{3/2}\sqrt{2}\,\mathrm{erf}\left(-\sqrt{2}\sqrt{a}\,L\right.$$

$$\left.+\sqrt{a}\left(x-\frac{px\,s2}{m}\right)\sqrt{2}\right)\mathrm{erf}\left(\sqrt{a}\left(y\right.\right.$$

$$\left.\left.-\frac{py\,s2}{m}\right)\sqrt{2}\right)e^{2a\left(x-\frac{px\,s2}{m}\right)^2+2aL^2}+\pi^{3/2}\sqrt{2}\,\mathrm{erf}\Bigg($$

$$-\sqrt{2}\sqrt{a}\,L+\sqrt{a}\left(x-\frac{px\,s2}{m}\right)\sqrt{2}\Bigg)\mathrm{erf}\left(\sqrt{2}\sqrt{a}\,L\right.$$

$$\left.-\sqrt{a}\left(y-\frac{py\,s2}{m}\right)\sqrt{2}\right)e^{2a\left(x-\frac{px\,s2}{m}\right)^2+2aL^2}\Bigg)\Bigg($$

$$-\mathrm{erf}\left(\sqrt{a}\left(z-\frac{pz\,s2}{m}\right)\sqrt{2}\right)+\mathrm{erf}\Bigg(-\sqrt{2}\sqrt{a}\,L+\sqrt{a}\left(z\right.$$

$$\frac{1}{32}\frac{1}{\sqrt{a}}\Bigg(g^2\Bigg(4\pi\left(x-\frac{px\,s2}{m}\right)\mathrm{erf}\left(\sqrt{a}\left(y-\frac{py\,s2}{m}\right)\sqrt{2}\right)\sqrt{a}\,e^{2aL^2}+4\pi\left(x-\frac{px\,s2}{m}\right)\mathrm{erf}\left(\sqrt{2}\sqrt{a}\,L-\sqrt{a}\left(y-\frac{py\,s2}{m}\right)\sqrt{2}\right)\sqrt{a}\,e^{2aL^2}-\pi^{3/2}\sqrt{2}\,\mathrm{erf}\left(\sqrt{a}\left(x-\frac{px\,s2}{m}\right)\sqrt{2}\right)\mathrm{erf}\left(\sqrt{a}\left(y-\frac{py\,s2}{m}\right)\sqrt{2}\right)e^{2a\left(x-\frac{px\,s2}{m}\right)^2+2aL^2}$$

$$-\pi^{3/2}\sqrt{2}\,\mathrm{erf}\left(\sqrt{a}\left(x-\frac{px\,s2}{m}\right)\sqrt{2}\right)\mathrm{erf}\left(\sqrt{2}\sqrt{a}\,L-\sqrt{a}\left(y-\frac{py\,s2}{m}\right)\sqrt{2}\right)e^{2a\left(x-\frac{px\,s2}{m}\right)^2+2aL^2}$$

$$+4\pi L\,e^{4aL\left(x-\frac{px\,s2}{m}\right)}\mathrm{erf}\left(\sqrt{a}\left(y-\frac{py\,s2}{m}\right)\sqrt{2}\right)\sqrt{a}$$

$$+4\pi L\,e^{4aL\left(x-\frac{px\,s2}{m}\right)}\mathrm{erf}\left(\sqrt{2}\sqrt{a}\,L-\sqrt{a}\left(y-\frac{py\,s2}{m}\right)\sqrt{2}\right)\sqrt{a}-4\pi\left(x-\frac{px\,s2}{m}\right)e^{4aL\left(x-\frac{px\,s2}{m}\right)}\mathrm{erf}\left(\sqrt{a}\left(y-\frac{py\,s2}{m}\right)\sqrt{2}\right)\sqrt{a}$$

$$-4\pi\left(x-\frac{px\,s2}{m}\right)e^{4aL\left(x-\frac{px\,s2}{m}\right)}\mathrm{erf}\left(\sqrt{2}\sqrt{a}\,L-\sqrt{a}\left(y-\frac{py\,s2}{m}\right)\sqrt{2}\right)\sqrt{a}+\pi^{3/2}\sqrt{2}\,\mathrm{erf}\left(-\sqrt{2}\sqrt{a}\,L+\sqrt{a}\left(x-\frac{px\,s2}{m}\right)\sqrt{2}\right)\mathrm{erf}\left(\sqrt{a}\left(y-\frac{py\,s2}{m}\right)\sqrt{2}\right)e^{2a\left(x-\frac{px\,s2}{m}\right)^2+2aL^2}+\pi^{3/2}\sqrt{2}\,\mathrm{erf}\left(-\sqrt{2}\sqrt{a}\,L+\sqrt{a}\left(x-\frac{px\,s2}{m}\right)\sqrt{2}\right)\mathrm{erf}\left(\sqrt{2}\sqrt{a}\,L-\sqrt{a}\left(y-\frac{py\,s2}{m}\right)\sqrt{2}\right)e^{2a\left(x-\frac{px\,s2}{m}\right)^2+2aL^2}\Bigg)\Bigg($$

$$-\mathrm{erf}\left(\sqrt{a}\left(z-\frac{pz\,s2}{m}\right)\sqrt{2}\right)+\mathrm{erf}\left(-\sqrt{2}\sqrt{a}\,L+\sqrt{a}\left(z\right.\right.$$

$$\frac{1}{32}\frac{1}{\sqrt{a}}\Bigg(g^2\Bigg(4\pi\left(x-\frac{px\,s2}{m}\right)\mathrm{erf}\left(\sqrt{a}\left(y\right.\right.$$

$$\left.\left.-\frac{py\,s2}{m}\right)\sqrt{2}\right)\sqrt{a}\ \mathrm{e}^{2aL^2}+4\pi\left(x\right.$$

$$\left.-\frac{px\,s2}{m}\right)\mathrm{erf}\left(\sqrt{2}\sqrt{a}\,L-\sqrt{a}\left(y\right.\right.$$

$$\left.\left.-\frac{py\,s2}{m}\right)\sqrt{2}\right)\sqrt{a}\ \mathrm{e}^{2aL^2}-\pi^{3/2}\sqrt{2}\ \mathrm{erf}\left(\sqrt{a}\left(x\right.\right.$$

$$\left.\left.-\frac{px\,s2}{m}\right)\sqrt{2}\right)\mathrm{erf}\left(\sqrt{a}\left(y\right.\right.$$

$$\left.\left.-\frac{py\,s2}{m}\right)\sqrt{2}\right)\mathrm{e}^{2a\left(x-\frac{px\,s2}{m}\right)^2+2aL^2}$$

$$-\pi^{3/2}\sqrt{2}\ \mathrm{erf}\left(\sqrt{a}\left(x-\frac{px\,s2}{m}\right)\sqrt{2}\right)\mathrm{erf}\left(\sqrt{2}\sqrt{a}\,L\right.$$

$$\left.-\sqrt{a}\left(y-\frac{py\,s2}{m}\right)\sqrt{2}\right)\mathrm{e}^{2a\left(x-\frac{px\,s2}{m}\right)^2+2aL^2}$$

$$+4\pi L\,\mathrm{e}^{4aL\left(x-\frac{px\,s2}{m}\right)}\mathrm{erf}\left(\sqrt{a}\left(y-\frac{py\,s2}{m}\right)\sqrt{2}\right)\sqrt{a}$$

$$+4\pi L\,\mathrm{e}^{4aL\left(x-\frac{px\,s2}{m}\right)}\mathrm{erf}\left(\sqrt{2}\sqrt{a}\,L-\sqrt{a}\left(y\right.\right.$$

$$\left.\left.-\frac{py\,s2}{m}\right)\sqrt{2}\right)\sqrt{a}-4\pi\left(x\right.$$

$$\left.-\frac{px\,s2}{m}\right)\mathrm{e}^{4aL\left(x-\frac{px\,s2}{m}\right)}\mathrm{erf}\left(\sqrt{a}\left(y-\frac{py\,s2}{m}\right)\sqrt{2}\right)\sqrt{a}$$

$$-4\pi\left(x-\frac{px\,s2}{m}\right)\mathrm{e}^{4aL\left(x-\frac{px\,s2}{m}\right)}\mathrm{erf}\left(\sqrt{2}\sqrt{a}\,L\right.$$

$$\left.-\sqrt{a}\left(y-\frac{py\,s2}{m}\right)\sqrt{2}\right)\sqrt{a}+\pi^{3/2}\sqrt{2}\ \mathrm{erf}\left(-\sqrt{2}\sqrt{a}\,L\right.$$

$$\left.+\sqrt{a}\left(x-\frac{px\,s2}{m}\right)\sqrt{2}\right)\mathrm{erf}\left(\sqrt{a}\left(y\right.\right.$$

$$\left.\left.-\frac{py\,s2}{m}\right)\sqrt{2}\right)\mathrm{e}^{2a\left(x-\frac{px\,s2}{m}\right)^2+2aL^2}+\pi^{3/2}\sqrt{2}\ \mathrm{erf}\Bigg($$

$$-\sqrt{2}\sqrt{a}\,L+\sqrt{a}\left(x-\frac{px\,s2}{m}\right)\sqrt{2}\Bigg)\mathrm{erf}\Bigg(\sqrt{2}\sqrt{a}\,L$$

$$-\sqrt{a}\left(y-\frac{py\,s2}{m}\right)\sqrt{2}\Bigg)\mathrm{e}^{2a\left(x-\frac{px\,s2}{m}\right)^2+2aL^2}\Bigg)\Bigg($$

$$-\mathrm{erf}\left(\sqrt{a}\left(z-\frac{pz\,s2}{m}\right)\sqrt{2}\right)+\mathrm{erf}\Bigg(-\sqrt{2}\sqrt{a}\,L+\sqrt{a}\left(z\right.$$

We have used the Maple code

Int3DVsqshifted := subs(x = x-px*s2/m, y = y-py*s2/m, z = z-pz*s2/m, Int3DVsq)

to produce (10).

Using (10), we perform the differentiations with respect to $x$.

All differentiation of shifted functions are done from right to left culminating in two iterated time integrations. The time integrations in $s_2$ and $s_1$ were done successively, and successfully, to give a very long expression for the time evolution of the x-momentum, for example.

The relevant Maple code executing all instructions are shown below:

```
> NULL;
>
> xmomentumpropagator := px*exp(-px*s2*Dx/m-py*s2*Dy/m-pz*s2*Dz/m);
> Dpx1momentumpropagator := diff(xmomentumpropagator, px);
> Dpx2momentumpropagator := diff(Dpx1momentumpropagator, px);
> xmomentumchange := Dpx2momentumpropagator*'Int3DVsq';
> xmomentumchange := (-2*s2*Dx/m+pz*s2^2*Dx^2/m^2)*'Int3DVsqshifted';
> Int3DVsqshifted;
>
>          xmomentumchange          :=          -2*s2*(diff(Int3DVsqshifted, x))/m+px*s2^2*(diff(diff(Int3DVsqshifted, x), x))/m^2;
>
> xmomentumchange := simplify(xmomentumchange);
> xmomentumchange := subs(py = 0, pz = 0, px = po, xmomentumchange);
```

>

> xmomentumchanges1 := int(xmomentumchange, s2 = 0 .. s1);

> xmomentumchanges1 := simplify(xmomentumchanges1);

> xmomentumchanget := int(xmomentumchanges1, s1 = 0 .. t);

> xmomentumchanget := simplify(xmomentumchanget);

> xmomentumchanget;

> xmom := po+xmomentumchanget;

> xmom;

to produce the following expression for the momentum in the x-direction:

(run Maple code in Appendix)
$$\tag{12}$$

Eq. (12) demonstrates that the time-dependent $x$- momentum is exact and analytic, a bit too long perhaps, but there is no other way of showing the solution for the $x$- momentum.

The symbolic code that caculates the $y$- momentum follows:

>ymomentumchange := -2*s2*(diff(Int3DVsqshifted, z))/m+py*s2^2*(diff(diff(Int3DVsqshifted, z), z))/m^2;

> ymomentumchange := simplify(ymomentumchange);

> ymomentumchange := subs(py = 0, pz = 0, px = po, ymomentumchange);

> ymomentumchanges1 := int(ymomentumchange, s2 = 0 .. s1);

> ymomentumchanges1 := simplify(ymomentumchanges1);

> ymomentumchanget := int(ymomentumchanges1, s1 = 0 .. t);

> ymomentumchanget := simplify(ymomentumchanget);

> ymom := ymomentumchanget;

to produce

(Run Maple code in Appendix)

We need not do the expression for *zmom*, the momentum in the *z*-direction because we only need to exchange *z*,*y* to get the analogue of (13) for the *z*- momentum. At this point, we have demonstrated that the analytic solutions exist and do not blow up in time. There also exist infinite time limits, which is best shown by time-dependent plots, not by repetitive and very long Maple outputs. Having demonstrated the existence of the solutions for the three momentum coordinates ad nauseaum, we plot the results using as examples the following Maple codes:

```
> with(plots);

> a := 'a'; g := 'g'; m := 'm'; L := 'L'; po := 'po';

>

> a := 1; g := 1; m := 1; L := 10; po := 10;

> xmom; xmom := simplify(xmom);

> y := 0; z := 0; T := 'T'; T := 2; plot3d(xmom, x = 0 .. 10, t = 0 .. T, axes = boxed, title = "x-momentum as a function of t"); T := 'T';

> zmom; zmom := simplify(zmom);

> y := 0; z := 0; T := 'T'; T := .2; plot3d(zmom, x = 0 .. 10, t = 0 .. T, axes = boxed, title = "z-momentum as a function of t"); T := 'T';

> y := 0; z := 0; T := 'T'; T := 1.0; plot3d(xmom, x = 0 .. 10, t = 0 .. T, axes = boxed, title = "x-momentum as a function of t"); T := 'T';

> y := 0; z := 0; T := 'T'; T := 2; plot3d(xmom, x = 0 .. 10, t = 0 .. T, axes = boxed, title = "x-momentum as a function of t"); T := 'T';

> zmom; zmom := simplify(zmom);

> y := 0; z := 0; T := 'T'; T := .3; plot3d(zmom, x = 0 .. 10, t = 0 .. T, axes = boxed, title = "z-momentum as a function of t"); T := 'T';
```

The results are shown in Figs. 1-4 in Cartesian form for the cube geometry.

We summarize the results graphically in two ways: Section 3 - Cartesian plots in a cube of dimension $L$, and Section - 4  3-d   toroid geometry by putting $x = L\sin(\theta), y = L\sin(\phi), z = L\sin(\omega)$.

5. Cartesian plots

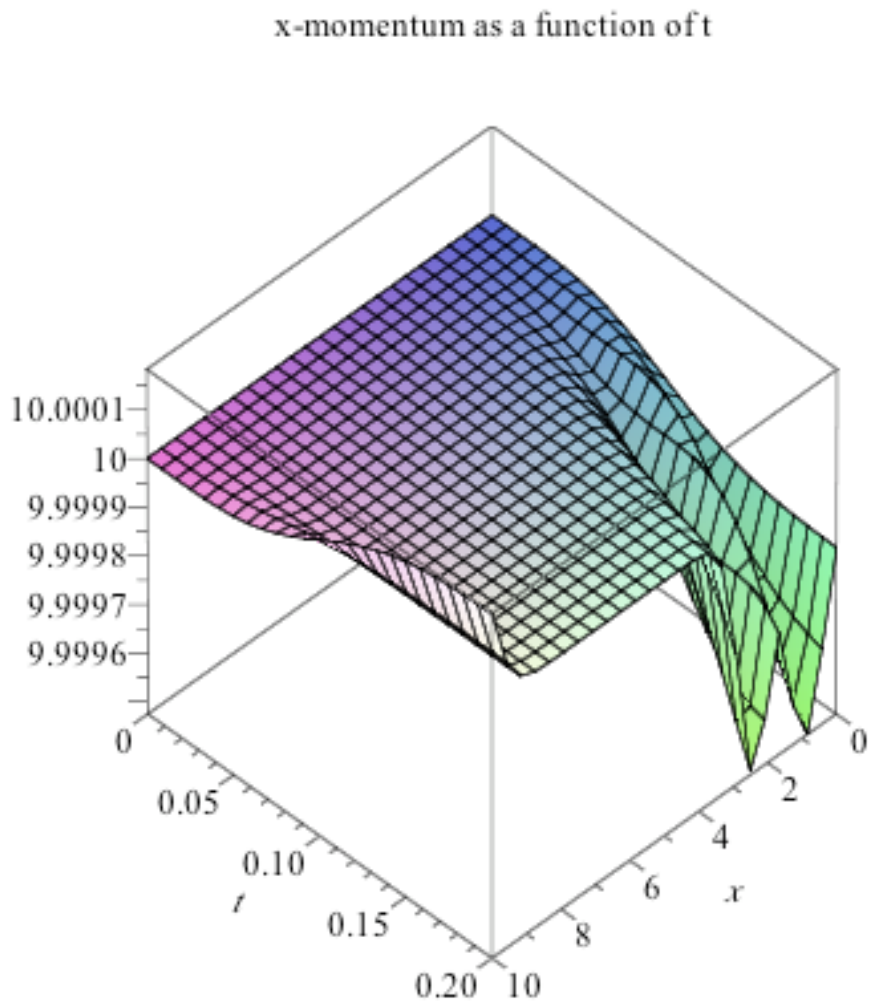

Fig. 1 – Cartesian plot: x-momentum at y, z=0,

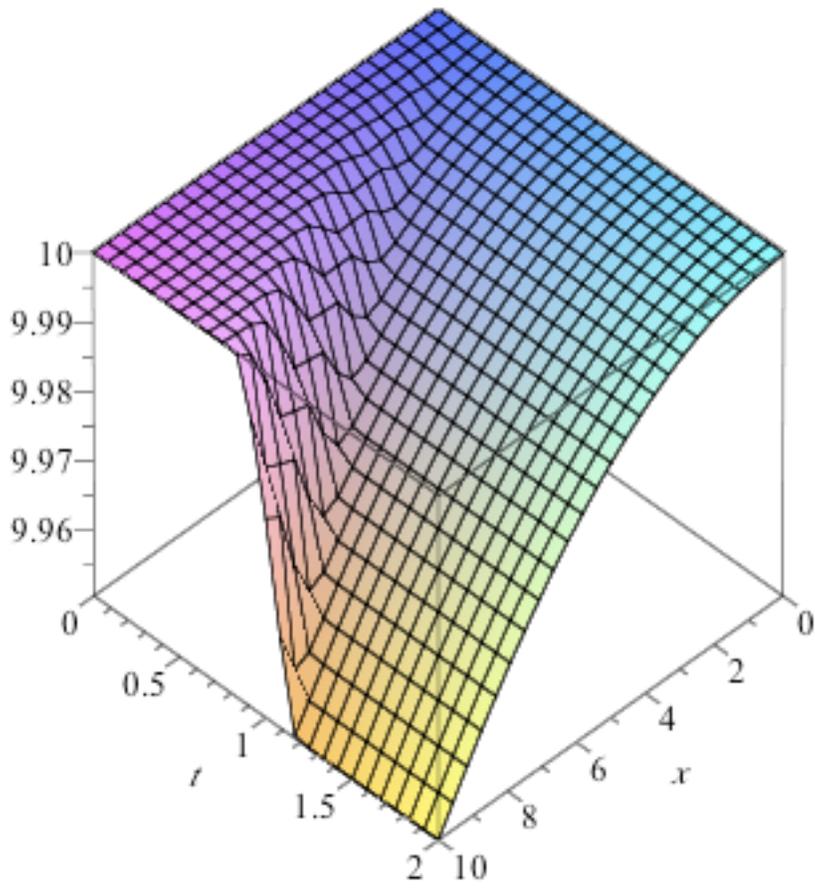

Fig. 2 – x-momentum for longer times.

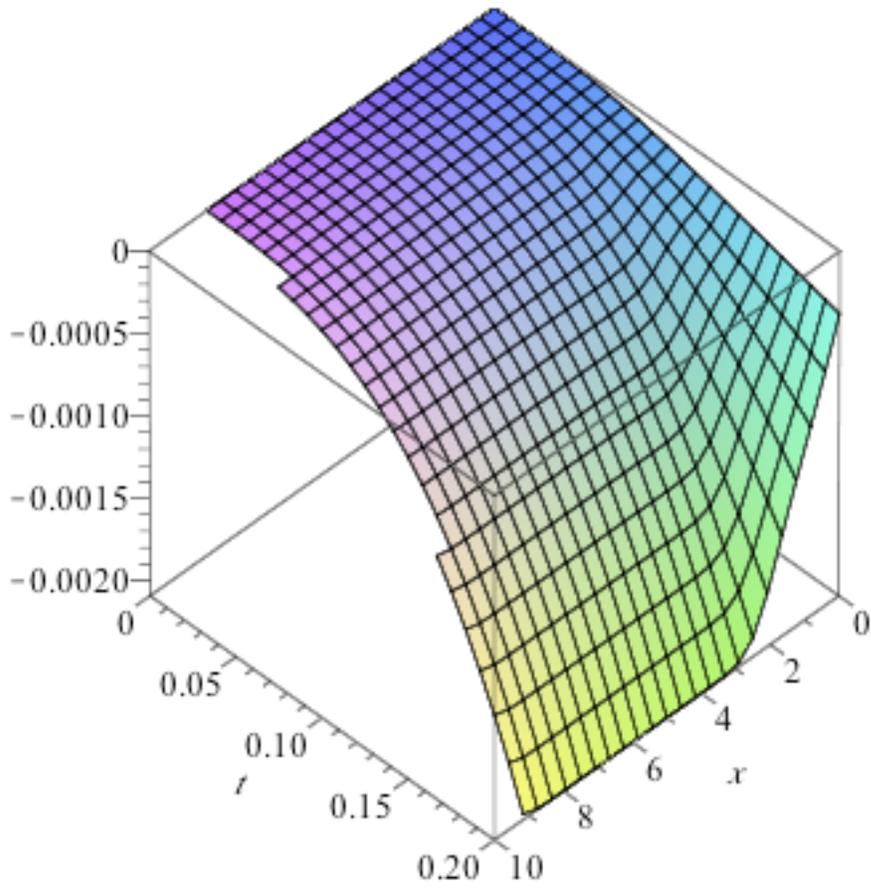

Fig. 3 – z-momentum as a function of time at y, z = 0.

## 6. 3-D Toroid plots

The following Maple code allows the transformation to 3-D toroid plots:

```
> NULL;

> x := 'x'; y := 'y'; z := 'z';

> theta := 'theta'; phi := 'phi'; omega := 'omega';

> a := 'a'; g := 'g'; m := 'm'; L := 'L'; po := 'po'; xmom; ymom; zmom;

> xtoroid := subs(x = L*sin(theta), y = L*sin(phi), z = L*sin(omega), xmom);

> ytoroid := subs(x = L*sin(theta), y = L*sin(phi), z = L*sin(omega), ymom);

> ztoroid := subs(x = L*sin(theta), y = L*sin(phi), z = L*sin(omega), zmom);

> a := 1; g := 1; m := 1; L := 10; po := 10;

> xtoroid; ytoroid; ztoroid;

> omega := 0;

> with(plots); t := 't'; t := 2.0; theta := 'theta'; phi := 'phi'; plot3d(xtoroid, theta = 0 .. Pi, phi = 0 .. Pi, axes = boxed, title = "x-momentum, omega=0, t = 2.0"); t := 't';

> with(plots); t := 't'; t := .1; theta := 'theta'; phi := 'phi'; plot3d(ztoroid, theta = 0 .. Pi, phi = 0 .. Pi, axes = boxed, title = "z-momentum, omega=0, t = 0.1"); t := 't';
```

We produce the following plots: Figs. 4 -

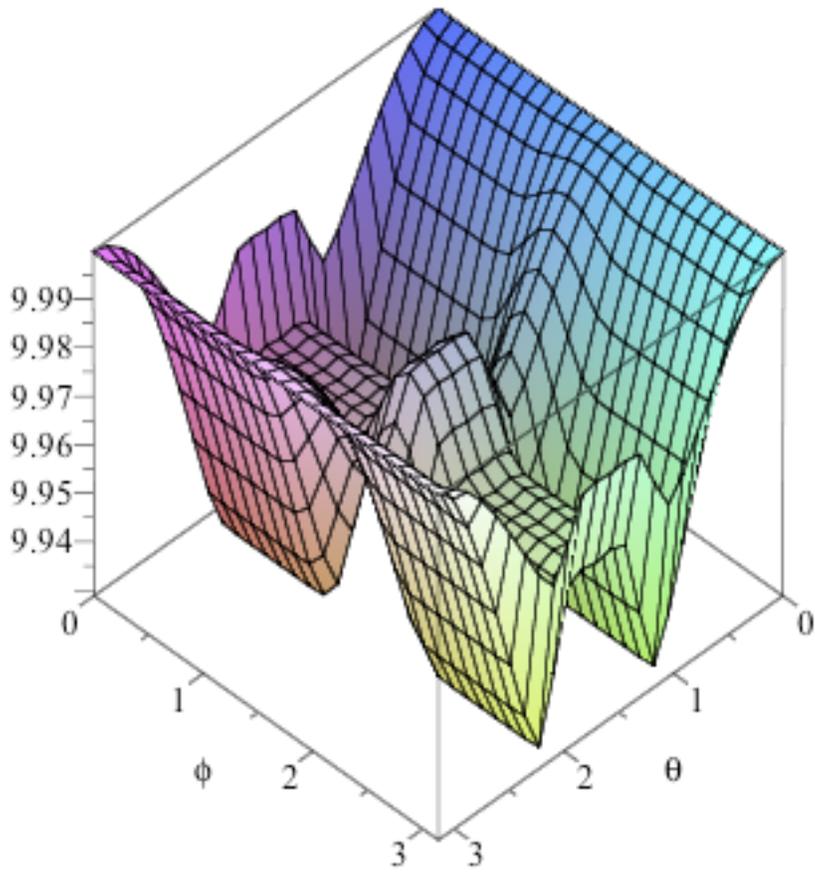

Fig. 4 – 3-d toroid plot of x – momentum as a function of theta and phi, omega = 0 , t = 0.1

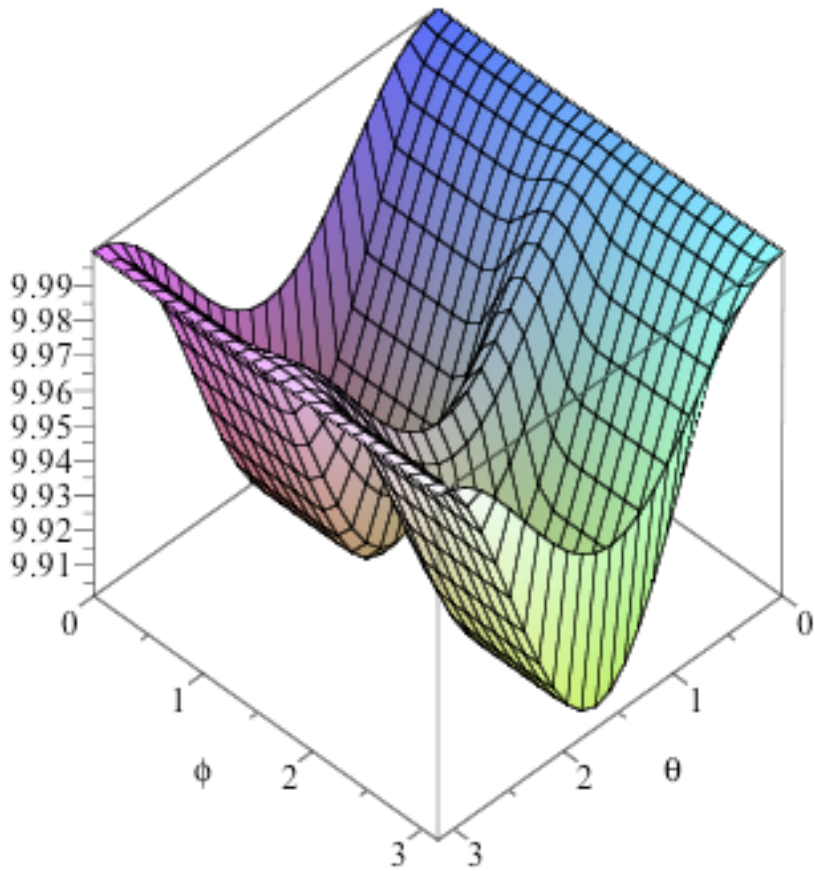

Fig. 5 – 3-d toroid plot of x – momentum as a function of theta and phi, omega = 0 , t = 2

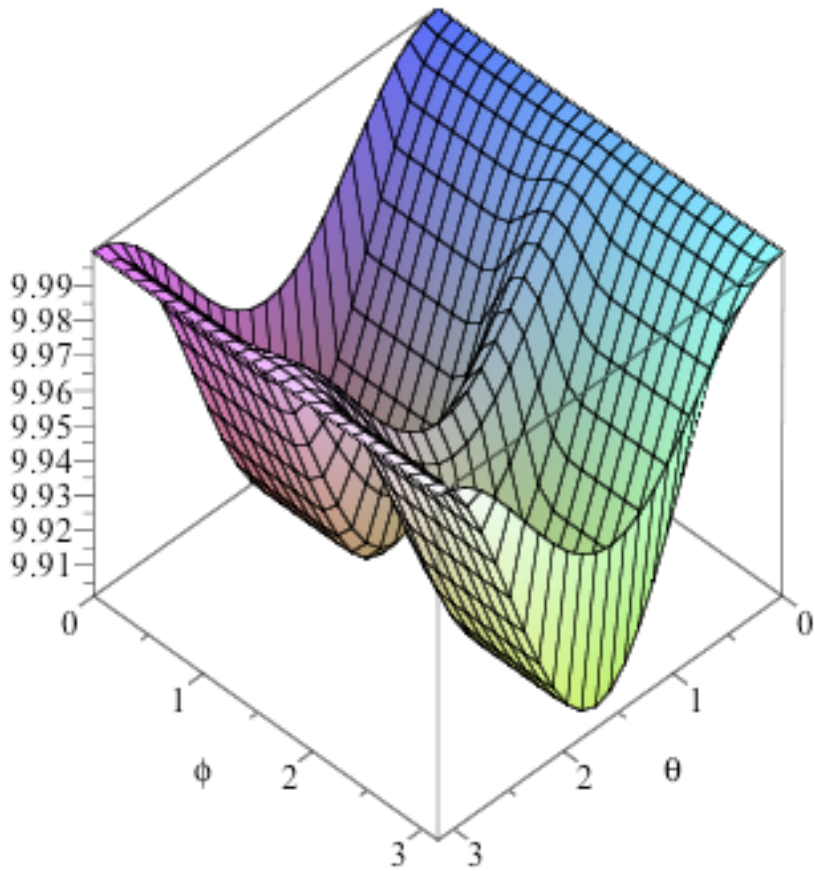

Fig. 6 – 3-d toroid plot of x – momentum as a function of theta and phi, omega = 0 , t = 200 illustrating long-time limit.

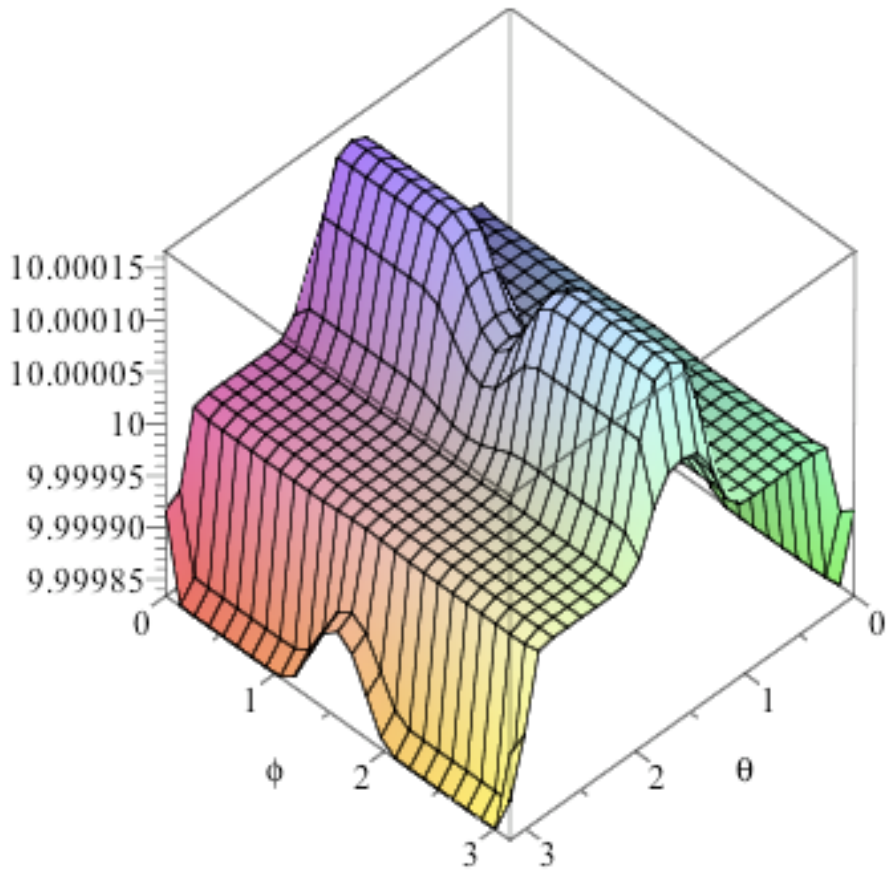

Fig. 7 – 3-d toroid plot of y-momentum as a function of theta and phi, omega = 0, t = 0.1

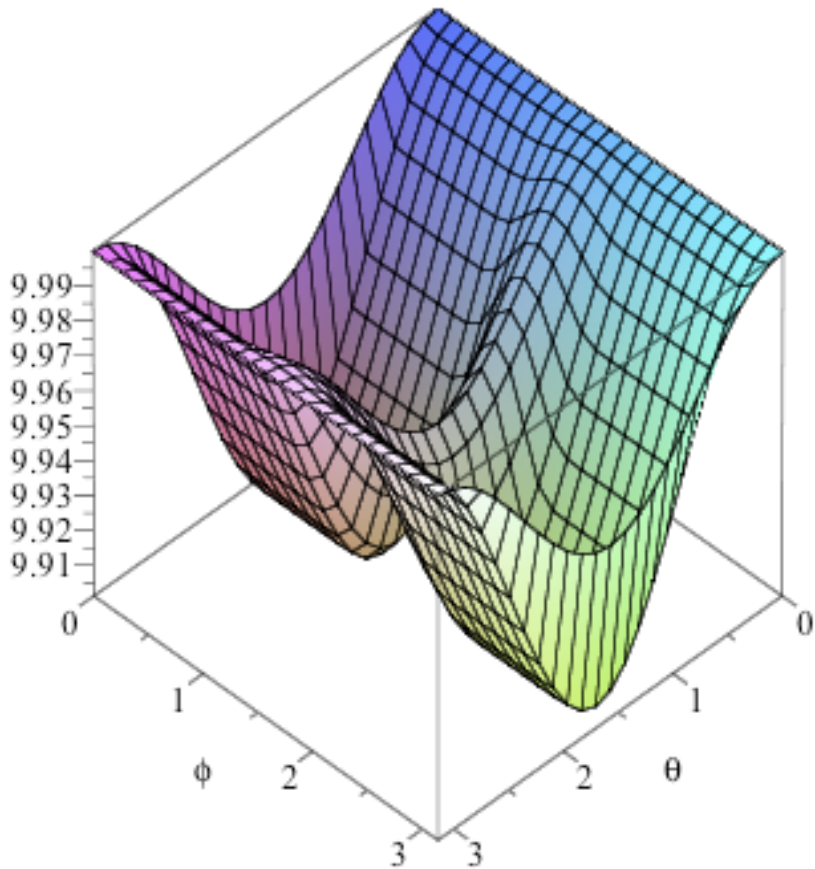

Fig. 8 – 3-d toroid plot of y-momentum as a function of theta and phi, omega = 0, t = 200, illustrating long-time limit.

5. Energy field

To calculate the energy field, we average $p_x^2/2m, p_y^2/2m, p_z^2/2m,$ the corresponding $G$ to be used from Eq. (8) is the following expression, an obvious preparation for Maple codes.

Dpx2energypropagator := 2*exp(-px*s2*Dx/m-py*s2*Dy/m-pz*s2*Dz/m)-4*px*s2*Dx*exp(-px*s2*Dx/m-py*s2*Dy/m-pz*s2*Dz/m)/m+px^2*s2^2*Dx^2*exp(-px*s2*Dx/m-py*s2*Dy/m-pz*s2*Dz/m)/m^2

(12)

again freely going from display functions to Maple code to show the harmonious use of display functions and Maple code. We simply repeat the procedure of Section to calculate the kinetic energy energy in the three Cartesian axes. For the kinetic energy in the x-direction we have the following code:

```
> NULL;

> NULL;

> xenergypropagator := px^2*exp(-px*s2*Dx/m-py*s2*Dy/m-pz*s2*Dz/m);

> Dpx1energypropagator := diff(xenergypropagator, px);

> Dpx2energypropagator := diff(Dpx1energypropagator, px);

> xenergychange := Dpx2energypropagator*'Int3DVsq';

> xenergychange := (2*exp(-px*s2*Dx/m-py*s2*Dy/m-pz*s2*Dz/m)-4*px*s2*Dx*exp(-px*s2*Dx/m-py*s2*Dy/m-pz*s2*Dz/m)/m+px^2*s2^2*Dx^2*exp(-px*s2*Dx/m-py*s2*Dy/m-pz*s2*Dz/m)/m^2)*'Int3DVsqshifted';

> Int3DVsqshifted;

> xenergychange := (2-4*px*s2*Dx/m+px^2*s2^2*Dx^2/m^2)*'Int3DVsqshifted';

> xenergychange := 2*Int3DVsqshifted-4*px*s2*(diff(Int3DVsqshifted, x))/m+px^2*s2^2*(diff(diff(Int3DVsqshifted, x), x))/m^2;

> xenergychange := simplify(xenergychange);

> xenergychange := subs(py = 0, pz = 0, px = po, xenergychange);

> xenergychanges1 := int(xenergychange, s2 = 0 .. s1);

> xenergychanges1 := simplify(xenergychanges1);

> xenergychanget := int(xenergychanges1, s1 = 0 .. t);

> xenergychanget := simplify(xenergychanget);
```

```
> xenergy := po^2/(2*m)+(1/2)*xenergychanget;

> xenergy := po^2/(2*m)+(1/2)*xenergychanget;

> xenergychanget := simplify(xenergychanget);

> xenergy := po^2/(2*m)+(1/2)*xenergychanget;
```

The Maple output for the kinetic energy in the x-direction is given by

xenergy := (1/2)*po^2/m+(1/256)*g^2*Pi*(erf(sqrt(a)*z*sqrt(2))+erf(sqrt(2)*sqrt(a)*(L-z)))*(-16*a^(3/2)*erf(sqrt(a)*y*sqrt(2))*exp(2*a*(po*t+L*m-x*m)^2/m^2)*x*m*t^2*po^2-16*a^(3/2)*erf(sqrt(2)*sqrt(a)*L-sqrt(a)*y*sqrt(2))*exp(2*a*(po*t+L*m-x*m)^2/m^2)*x*m*t^2*po^2+16*a^(3/2)*erf(sqrt(a)*y*sqrt(2))*exp(2*a*(-po*t+x*m)^2/m^2)*m*x*t^2*po^2-16*a^(3/2)*erf(sqrt(a)*y*sqrt(2))*exp(2*a*(-po*t+x*m)^2/m^2)*m*L*t^2*po^2+16*a^(3/2)*erf(sqrt(2)*sqrt(a)*L-sqrt(a)*y*sqrt(2))*exp(2*a*(-po*t+x*m)^2/m^2)*m*x*t^2*po^2-16*a^(3/2)*erf(sqrt(2)*sqrt(a)*L-sqrt(a)*y*sqrt(2))*exp(2*a*(-po*t+x*m)^2/m^2)*m*L*t^2*po^2-m^3*erf(sqrt(2)*sqrt(a)*L-sqrt(a)*y*sqrt(2))*sqrt(Pi)*sqrt(2)*erf(sqrt(2)*sqrt(a)*(-po*t+x*m)/m)*exp(2*a*(2*x^2*m^2-4*m*po*t*x+2*po^2*t^2+2*m*po*t*L+L^2*m^2-2*L*x*m^2)/m^2)+4*sqrt(a)*erf(sqrt(2)*sqrt(a)*L-sqrt(a)*y*sqrt(2))*exp(2*a*(po*t+L*m-x*m)^2/m^2)*m^2*t*po+4*m^2*erf(sqrt(2)*sqrt(a)*(L-y))*sqrt(a)*exp(2*a*(-po*t+x*m)^2/m^2)*po*t-4*m^2*erf(sqrt(2)*sqrt(a)*(L-y))*sqrt(a)*exp(2*a*(po*t+L*m-x*m)^2/m^2)*po*t+m^3*erf(sqrt(2)*sqrt(a)*(L-y))*sqrt(Pi)*sqrt(2)*erf(sqrt(2)*sqrt(a)*(po*t+L*m-x*m)/m)*exp(2*a*(2*x^2*m^2-4*m*po*t*x+2*po^2*t^2+2*m*po*t*L+L^2*m^2-2*L*x*m^2)/m^2)-m^3*erf(sqrt(2)*sqrt(a)*L-sqrt(a)*y*sqrt(2))*sqrt(Pi)*sqrt(2)*erf(sqrt(2)*sqrt(a)*(po*t+L*m-x*m)/m)*exp(2*a*(2*x^2*m^2-4*m*po*t*x+2*po^2*t^2+2*m*po*t*L+L^2*m^2-2*L*x*m^2)/m^2)+m^3*erf(sqrt(2)*sqrt(a)*(L-y))*sqrt(Pi)*sqrt(2)*erf(sqrt(2)*sqrt(a)*(-po*t+x*m)/m)*exp(2*a*(2*x^2*m^2-4*m*po*t*x+2*po^2*t^2+2*m*po*t*L+L^2*m^2-2*L*x*m^2)/m^2)-4*sqrt(a)*erf(sqrt(2)*sqrt(a)*L-sqrt(a)*y*sqrt(2))*exp(2*a*(-po*t+x*m)^2/m^2)*m^2*t*po-8*m^2*a*erf(sqrt(2)*sqrt(a)*(L-y))*x*exp(2*a*(2*x^2*m^2-4*m*po*t*x+2*po^2*t^2+2*m*po*t*L+L^2*m^2-2*L*x*m^2)/m^2)*erf(sqrt(2)*sqrt(a)*(po*t+L*m-x*m)/m)*sqrt(2)*sqrt(Pi)*po*t+4*m*sqrt(2)*erf(sqrt(a)*y*sqrt(2))*a*exp(2*a*(2*x^2*m^2-4*m*po*t*x+2*po^2*t^2+2*m*po*t*L+L^2*m^2-2*L*x*m^2)/m^2)*erf(sqrt(2)*sqrt(a)*(-po*t+x*m)/m)*sqrt(Pi)*po^2*t^2+8*m^3*x*erf(sqrt(2)*sqrt(a)*L-sqrt(a)*y*sqrt(2))*a*exp(2*a*(2*x^2*m^2-4*m*po*t*x+2*po^2*t^2+2*m*po*t*L+L^2*m^2-2*L*x*m^2)/m^2)*erf(sqrt(2)*sqrt(a)*(po*t+L*m-x*m)/m)*sqrt(2)*sqrt(Pi)*L+8*m^2*a*erf(sqrt(2)*sqrt(a)*(L-y))*L*exp(2*a*(2*x^2*m^2-

4*m*po*t*x+2*po^2*t^2+2*m*po*t*L+L^2*m^2-2*L*x*m^2)/m^2)*erf(sqrt(2)*sqrt(a)*(po*t+L*m-x*m)/m)*sqrt(2)*sqrt(Pi)*po*t-8*m^3*a*erf(sqrt(2)*sqrt(a)*(L-y))*L*exp(2*a*(2*x^2*m^2-4*m*po*t*x+2*po^2*t^2+2*m*po*t*L+L^2*m^2-2*L*x*m^2)/m^2)*erf(sqrt(2)*sqrt(a)*(po*t+L*m-x*m)/m)*sqrt(2)*sqrt(Pi)*x-8*m^2*a*erf(sqrt(2)*sqrt(a)*(L-y))*x*exp(2*a*(2*x^2*m^2-4*m*po*t*x+2*po^2*t^2+2*m*po*t*L+L^2*m^2-2*L*x*m^2)/m^2)*erf(sqrt(2)*sqrt(a)*(-po*t+x*m)/m)*sqrt(2)*sqrt(Pi)*po*t+4*m*sqrt(2)*erf(sqrt(2)*sqrt(a)*(L-y))*a*exp(2*a*(2*x^2*m^2-4*m*po*t*x+2*po^2*t^2+2*m*po*t*L+L^2*m^2-2*L*x*m^2)/m^2)*erf(sqrt(2)*sqrt(a)*(-po*t+x*m)/m)*sqrt(Pi)*po^2*t^2+4*m*sqrt(2)*erf(sqrt(a)*y*sqrt(2))*a*exp(2*a*(2*x^2*m^2-4*m*po*t*x+2*po^2*t^2+2*m*po*t*L+L^2*m^2-2*L*x*m^2)/m^2)*erf(sqrt(2)*sqrt(a)*(po*t+L*m-x*m)/m)*sqrt(Pi)*po^2*t^2+8*m^2*x*erf(sqrt(2)*sqrt(a)*L-sqrt(a)*y*sqrt(2))*a*exp(2*a*(2*x^2*m^2-4*m*po*t*x+2*po^2*t^2+2*m*po*t*L+L^2*m^2-2*L*x*m^2)/m^2)*erf(sqrt(2)*sqrt(a)*(po*t+L*m-x*m)/m)*sqrt(2)*sqrt(Pi)*po*t+4*m*sqrt(2)*erf(sqrt(2)*sqrt(a)*(L-y))*a*exp(2*a*(2*x^2*m^2-4*m*po*t*x+2*po^2*t^2+2*m*po*t*L+L^2*m^2-2*L*x*m^2)/m^2)*erf(sqrt(2)*sqrt(a)*(po*t+L*m-x*m)/m)*sqrt(Pi)*po^2*t^2+8*m^2*x*erf(sqrt(2)*sqrt(a)*L-sqrt(a)*y*sqrt(2))*a*exp(2*a*(2*x^2*m^2-4*m*po*t*x+2*po^2*t^2+2*m*po*t*L+L^2*m^2-2*L*x*m^2)/m^2)*erf(sqrt(2)*sqrt(a)*(-po*t+x*m)/m)*sqrt(2)*sqrt(Pi)*po*t-4*m^3*x*erf(sqrt(2)*sqrt(a)*L-sqrt(a)*y*sqrt(2))*sqrt(a)*exp(2*a*(po*t+L*m-x*m)^2/m^2)-4*m^3*erf(sqrt(2)*sqrt(a)*L-sqrt(a)*y*sqrt(2))*L*sqrt(a)*exp(2*a*(-po*t+x*m)^2/m^2)+4*m^3*erf(sqrt(2)*sqrt(a)*(L-y))*sqrt(a)*exp(2*a*(-po*t+x*m)^2/m^2)*L-4*m^3*erf(sqrt(2)*sqrt(a)*(L-y))*sqrt(a)*exp(2*a*(-po*t+x*m)^2/m^2)*x+4*m^3*erf(sqrt(2)*sqrt(a)*(L-y))*sqrt(a)*exp(2*a*(po*t+L*m-x*m)^2/m^2)*x+4*m^3*x*erf(sqrt(2)*sqrt(a)*L-sqrt(a)*y*sqrt(2))*sqrt(a)*exp(2*a*(-po*t+x*m)^2/m^2)-16*a^(3/2)*erf(sqrt(2)*sqrt(a)*L-sqrt(a)*y*sqrt(2))*exp(2*a*(-po*t+x*m)^2/m^2)*t^3*po^3-16*a^(3/2)*erf(sqrt(a)*y*sqrt(2))*exp(2*a*(-po*t+x*m)^2/m^2)*t^3*po^3+16*a^(3/2)*erf(sqrt(2)*sqrt(a)*L-sqrt(a)*y*sqrt(2))*exp(2*a*(po*t+L*m-x*m)^2/m^2)*t^3*po^3+16*a^(3/2)*erf(sqrt(a)*y*sqrt(2))*exp(2*a*(po*t+L*m-x*m)^2/m^2)*t^3*po^3-8*m^2*erf(sqrt(2)*sqrt(a)*L-sqrt(a)*y*sqrt(2))*L*a*exp(2*a*(2*x^2*m^2-4*m*po*t*x+2*po^2*t^2+2*m*po*t*L+L^2*m^2-2*L*x*m^2)/m^2)*erf(sqrt(2)*sqrt(a)*(po*t+L*m-x*m)/m)*sqrt(2)*sqrt(Pi)*po*t+4*m^3*a*erf(sqrt(2)*sqrt(a)*(L-y))*x^2*exp(2*a*(2*x^2*m^2-4*m*po*t*x+2*po^2*t^2+2*m*po*t*L+L^2*m^2-2*L*x*m^2)/m^2)*erf(sqrt(2)*sqrt(a)*(po*t+L*m-x*m)/m)*sqrt(2)*sqrt(Pi)-

4*a*erf(sqrt(2)*sqrt(a)*L-sqrt(a)*y*sqrt(2))*exp(2*a*(2*x^2*m^2-4*m*po*t*x+2*po^2*t^2+2*m*po*t*L+L^2*m^2-2*L*x*m^2)/m^2)*x^2*m^3*erf(sqrt(2)*sqrt(a)*(po*t+L*m-x*m)/m)*sqrt(2)*sqrt(Pi)-4*a*erf(sqrt(2)*sqrt(a)*L-sqrt(a)*y*sqrt(2))*exp(2*a*(2*x^2*m^2-4*m*po*t*x+2*po^2*t^2+2*m*po*t*L+L^2*m^2-2*L*x*m^2)/m^2)*m^3*sqrt(Pi)*sqrt(2)*erf(sqrt(2)*sqrt(a)*(po*t+L*m-x*m)/m)*L^2-4*a*erf(sqrt(2)*sqrt(a)*L-sqrt(a)*y*sqrt(2))*exp(2*a*(2*x^2*m^2-4*m*po*t*x+2*po^2*t^2+2*m*po*t*L+L^2*m^2-2*L*x*m^2)/m^2)*x^2*m^3*sqrt(Pi)*sqrt(2)*erf(sqrt(2)*sqrt(a)*(-po*t+x*m)/m)+4*m^3*a*erf(sqrt(2)*sqrt(a)*(L-y))*L^2*exp(2*a*(2*x^2*m^2-4*m*po*t*x+2*po^2*t^2+2*m*po*t*L+L^2*m^2-2*L*x*m^2)/m^2)*erf(sqrt(2)*sqrt(a)*(po*t+L*m-x*m)/m)*sqrt(2)*sqrt(Pi)+4*m^3*a*erf(sqrt(2)*sqrt(a)*(L-y))*x^2*exp(2*a*(2*x^2*m^2-4*m*po*t*x+2*po^2*t^2+2*m*po*t*L+L^2*m^2-2*L*x*m^2)/m^2)*erf(sqrt(2)*sqrt(a)*(-po*t+x*m)/m)*sqrt(2)*sqrt(Pi))*exp(-2*a*(2*x^2*m^2-4*m*po*t*x+2*po^2*t^2+2*m*po*t*L+L^2*m^2-2*L*x*m^2)/m^2)/(a^(3/2)*m*po^2)

(13)

The results for the time evolution of the energy density in the x, y, z –direction are shown in the plots of Figs. 9 – 11.

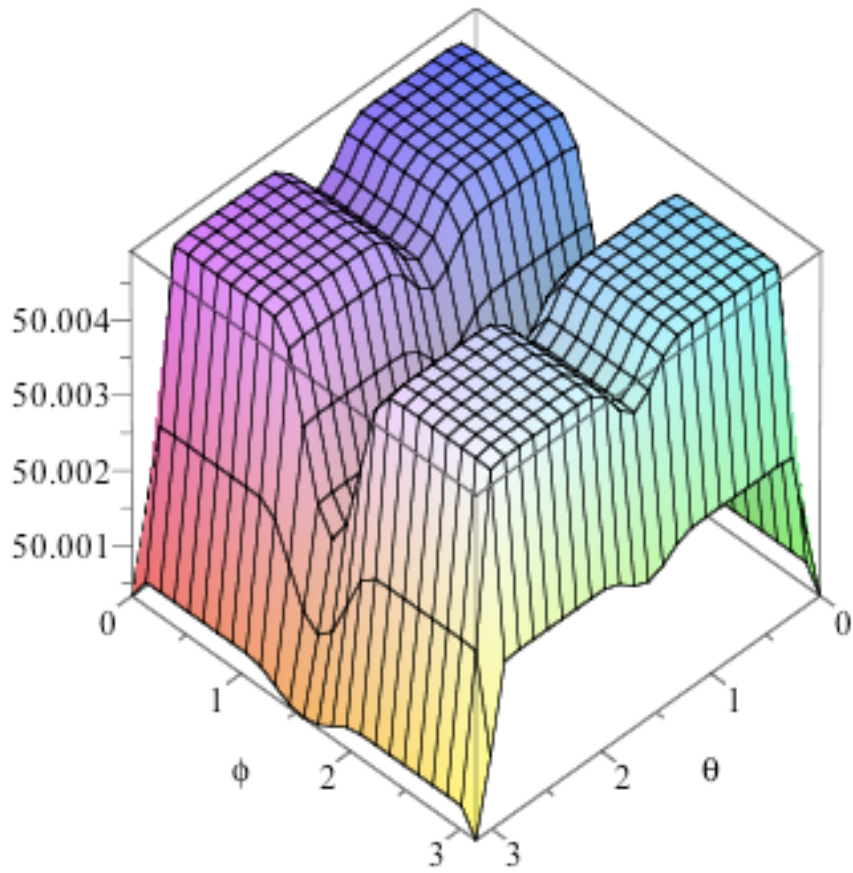

Fig. 9 – 3-d toroid plot: kinetic energy density in x as a function of theta, phi, omega = 0, t = 0.1

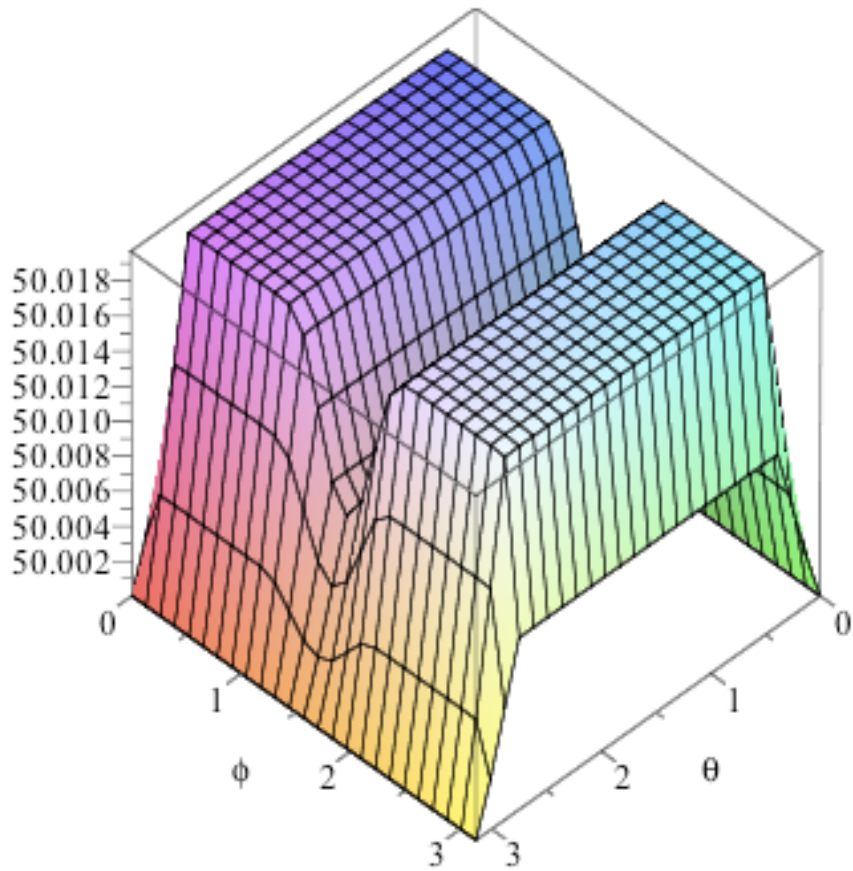

Fig. 10 – 3-d toroid plot: kinetic energy density in x as a function of theta, phi, omega = 0, t = 0.2

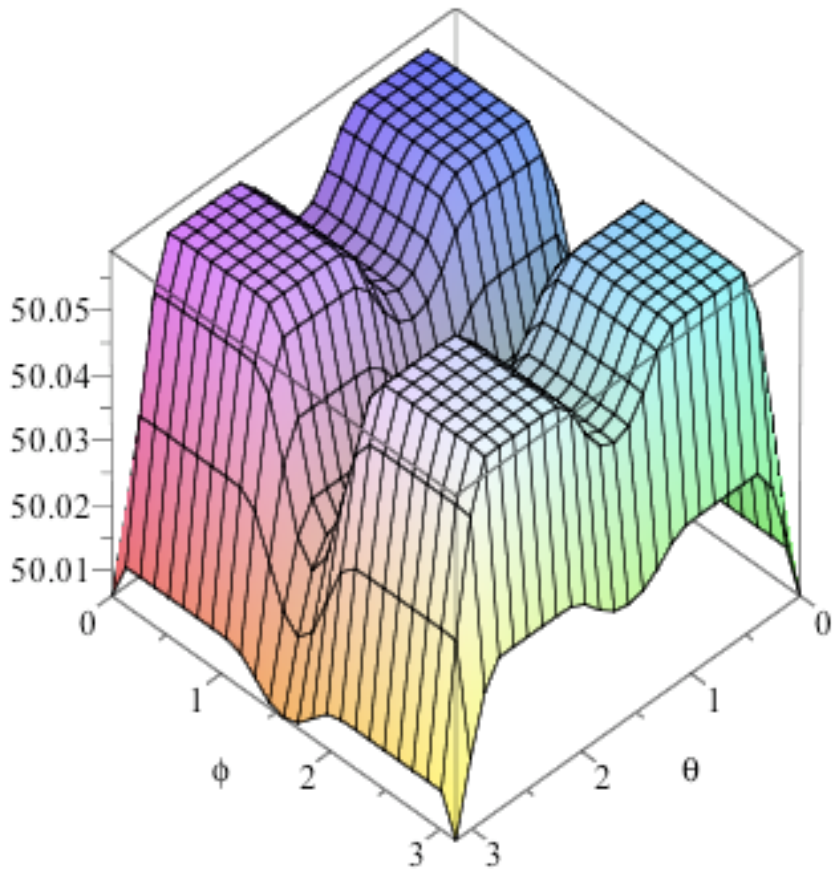

Fig. 11 – 3-d-toroid plot of total kinetic energy as a function of theta, phi, omega = 0, t = 0.2

### 7. Pressure calculations

Our time evolution results for the energy density, by virtue of its form, also yield the pressure from an elementary definition of pressure as the transport of momentum given by

$$P_\alpha(x,y,z,t) = 2\int dp_\alpha p_\alpha^2 f(x,p,z,t) \tag{14}$$

which is 4 times the energy density expressions for $m=1$. All other adjustable parameters are unity, which we use throughout all plots.

Long time limits exist for all velocity and energy fields, as well as the pressure. They may be written down, but again, even after reducing the number of expressions, the terms are still forbidding in number, one may just as well increase time, to say t = 200, and obtain the limiting plots at t = infinity. In the future, we hope to address the fundamental question of steady states and irreversibility, yet another paradox in non-equilibrium statistical mechanics.

8.  Divergence of the velocity field

One requirement of the Clay problem definition is that $u = p/m$ must satisfy

$$\nabla \cdot u = 0 \tag{15}$$

to show incompressibility. From our TE solutions, this is a mathematical requirement, but not a physical one. Our assumption of initial uniformity is sufficient. Nevertheless, we fed the left hand side of this equation the solutions from TE, hoping that the Maple code will show this analytically and rigorously, but the sheer number of expressions is quite forbidding, so we generated instead Fig. 14, showing that in fact, for most of the volume of the cube, the divergence is zero. At the boundaries, the condition no longer holds.

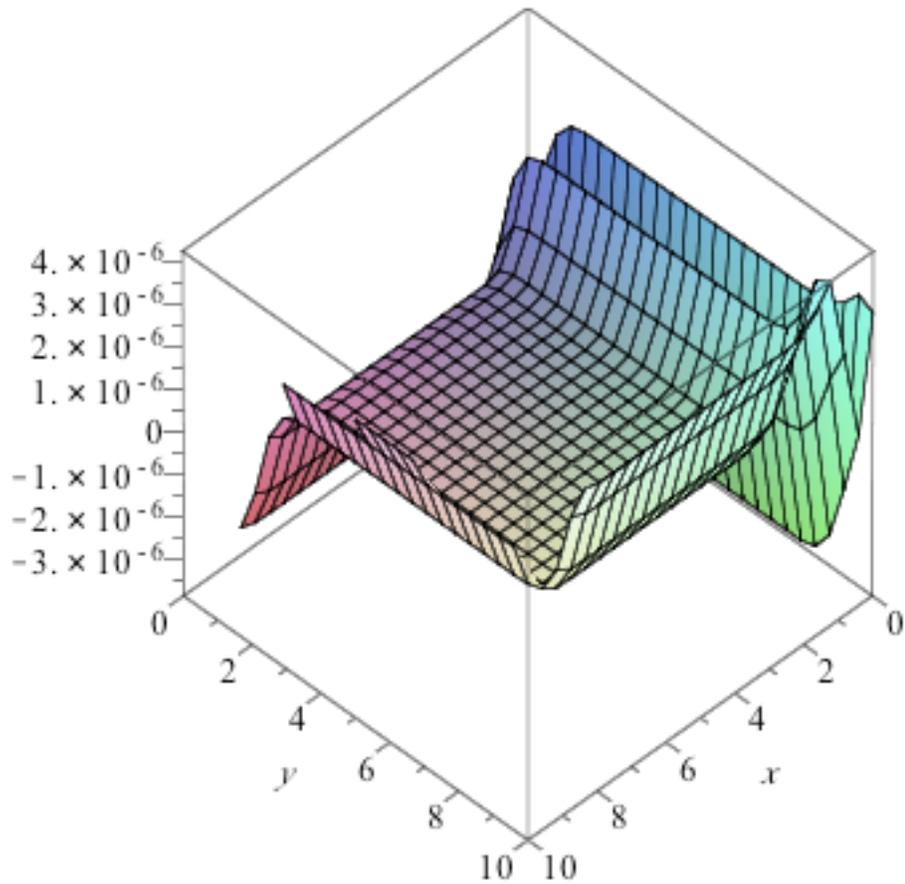

Fig. 14 – Divergence of the velocity at the plane $z = 0$ and $t = 0.02$. The plateau shows zero divergence.

Thus our time evolution equations yield the velocity fields, energy fields and pressure. We do not need to limit ourselves to incompressible systems as specified by the Clay requirements, but we follow the divergence stipulation automatically from the TE solutions, except at the boundaries of the cube in the Cartesian plots. The self-consistent non-linear solutions all satisfy the Clay stipulations, continuous fields and no blowup times. But the solutions do not come from a frontal attack on NSE directly. By the mere fact that our solutions possess "higher pedigree", coming from direct contracted results from the Liouville equation, we trust our time evolution solutions more. They reflect a more fundamental approach. NSE, by contrast, is a contraction from various kinetic equations, which by themselves are approximations. The continuum model adopted by NSE ignore molecular physics. NSE was proposed long before atoms and molecules were discovered. It is amazing that NSE has ruled the attempts to study turbulence ever since, at great expense and effort, not to speak of the fruitlessness in explaining turbulence. We will discuss this point later.

9. Exact solution of Navier-Stokes equation

Despite our comments about the provenance of solutions by the time evolution equations (TE) we still wish to address the problem of solving NSE. We can now substitute the velocity fields obtained from the time evolution equations to calculate from NSE the corresponding expression $DPx$ in our Maple codes, the derivative of pressure with respect to x, from the following form of NSE, written with a new philosophical twist, for example

$$\frac{\partial P}{\partial x} = \nu \nabla^2 u_x - u \bullet \nabla u_x - \frac{\partial u_x}{\partial t}$$

(16)

where we now use the velocity $u = \frac{p}{m}, m = 1$ instead of $p$.

Admittedly no one seems to write the Navier-Stokes Equation in the form of Eq. (16) for the x-component of the velocity, say, but we have all expressions for the velocities, and what we now do is evaluate the right-hand side of NSE in Eq. (16) and find the gradient of the pressure in the x-direction, as above, for example.

Using previous Maple outputs for $xmom, ymom, zmom$, the momentum in Cartesian coordinates,

```
> Dxmomdt := diff(xmom, t);
> Dxmomdx := diff(xmom, x);
> Dxmomdx2 := diff(Dxmomdx, x);
> Dxmomdy := diff(xmom, y);
> Dxmomdy2 := diff(Dxmomdy, x);
> Dxmomdz := diff(xmom, z);
> Dxmomdz2 := diff(Dxmomdz, z);
> gradx := xmom·Dxmomdx + ymom·Dxmomdy + zmom·Dxmomdz;
> xnabla := ν·Dxmomdx2 + ν·Dxmomdy2 + ν·Dxmomdz2;
```

We calculate the $x$-gradient of the pressure or force per unit area in the x-direction. In the above notation, for self-documentation, *xmom* refers to the momentum in the $x$-direction, *ymom* to the momentum in the $y$-direction and *zmom* the momentum in the $z$-direction. With the above notation we see the full force of the non-linearity of NSE.

From the above, we solve for the gradient of the pressure in the $x$-direction from NSE:

$$DPx := xnabla - gradx - Dxmomdt ; \qquad (17)$$

which is simply NSE written in a form that reflects the above-mentioned philosophical shift.

Again, we can write the full analytic expressions, including some hundreds of terms. If needed the above programs may be executed in a laptop with very little power.

We plot the $x$-gradient of the pressure as a function of time for the 3-d toroid geometry for two times using NSE

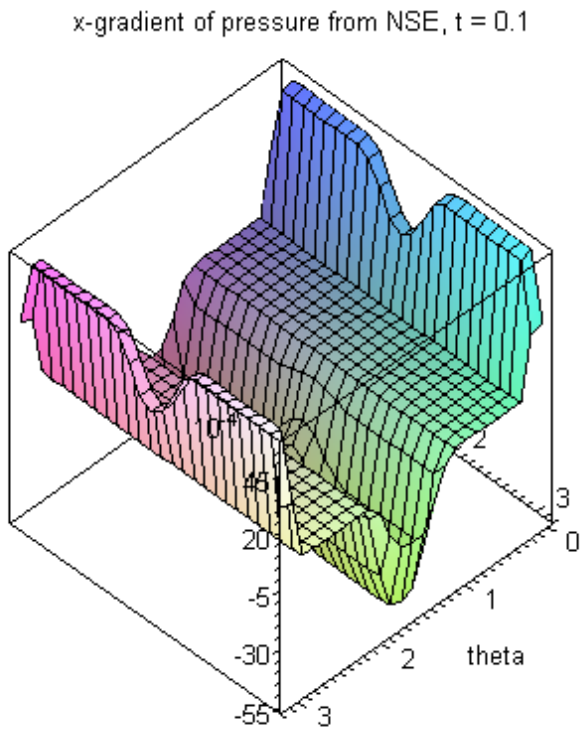

Fig. 14 -3d toroid plot of x-gradient of pressure as a function of theta, phi, omega = 0 at t 0.1

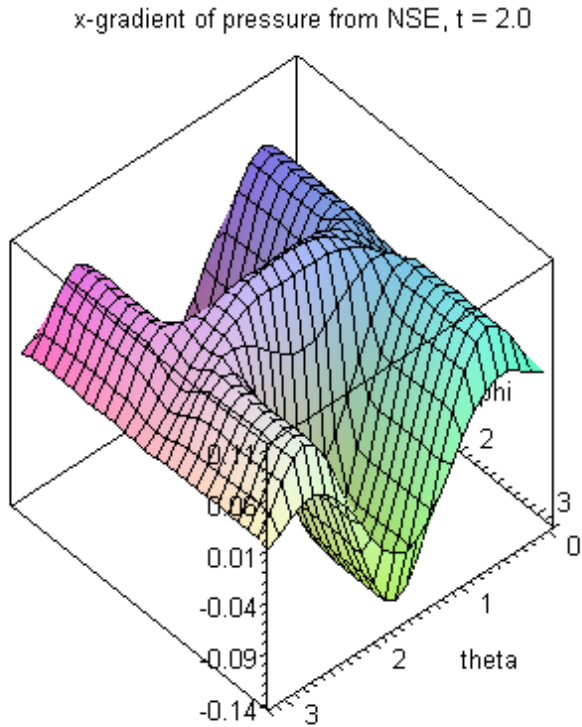

Fig. 15 – 3d toroid plot of x-gradient of pressure as a function of theta, phi, omega = 0 at t = 2

**10.   Comparison of pressure as calculated from TE and NSE.**

   We show the difference in approach by calculating the $x$-gradient of the pressure in the $x$-direction from NSE and the derivative of the pressure obtained from the TE solution . In NSE, we have expressions for the derivative of pressure with respect to coordinates, available as well in symbolic code. The pressure itself may be obtained by integration, symbolically, or numerically. But the symbolic integration cannot be completed by Maple beyond a certain point. So we calculated the derivative from NSE, and independently, differentiate our result for pressure from TE, (An old amusing adage from Uhlenbeck, "It is simpler to differentiate than to integrate."),  The results are qualitatively the same, but we prefer the TE result, by reason of the limited fundamental basis for NSE.  The plots are in Fig. 17 and Fig. 18.

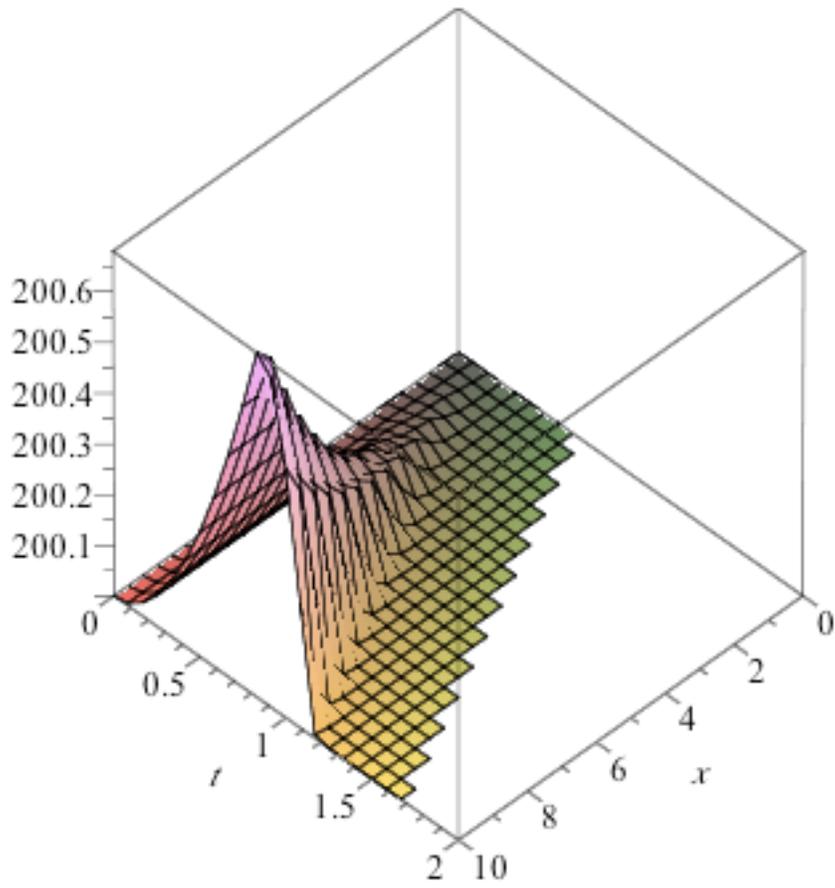

Fig. 16 – Time evolution of x- pressure from TE.

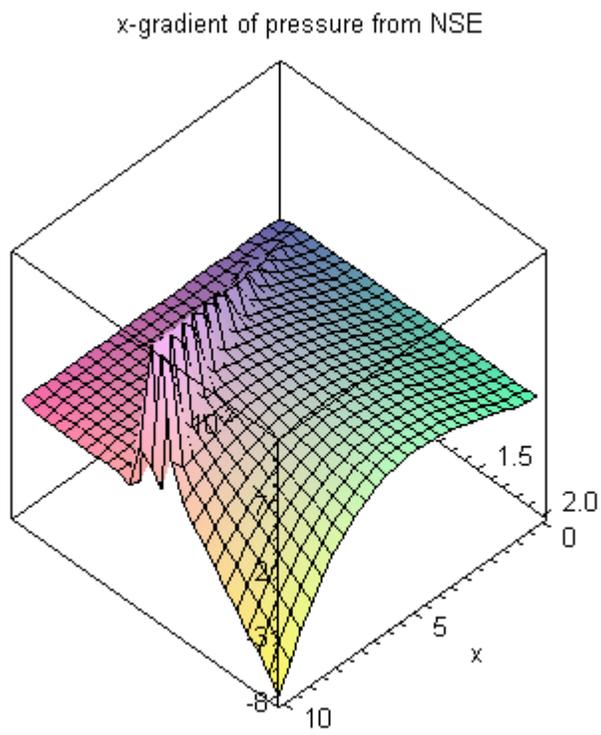

Fig. 17 –x-gradient of pressure as a function of time from NSE.

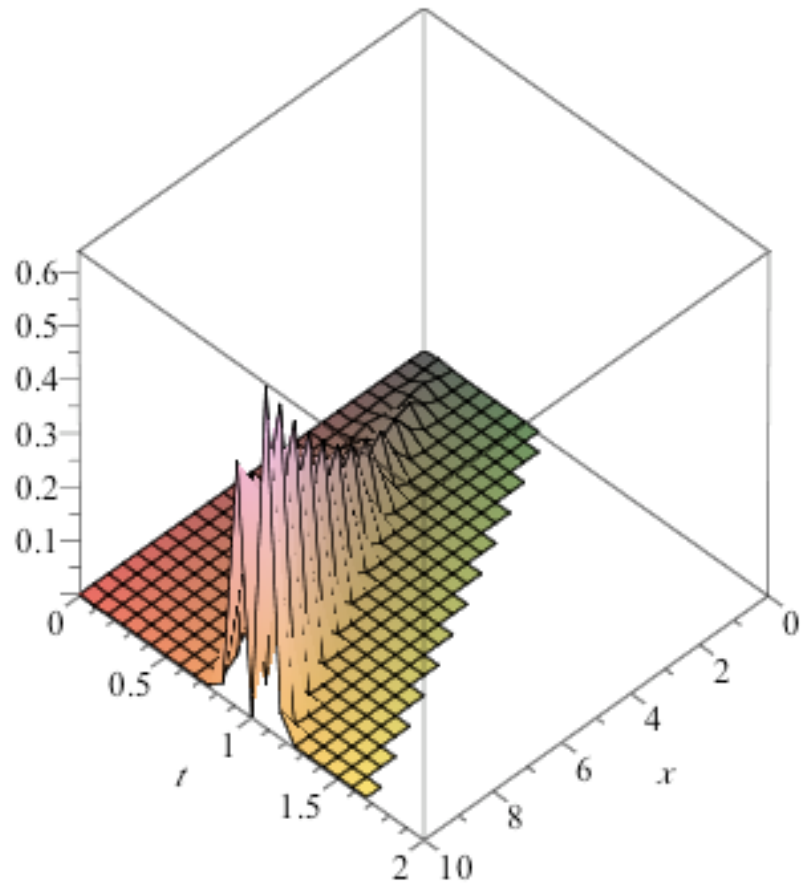

**Fig. 18 - x-gradient of x-pressure from TE.**

Both $y, z$ are put to zero, employing the same adjustable constants used in all plots (unity). As $t$ goes to infinity, the gradient of pressure from TE goes to zero. The pressure is well-behaved for various times from the energy plots. Why do we show the gradient of the pressure as calculated from TE? The pressure itself in the x-direction may be found by simply integrating the x-gradient of the pressure as calculated from NSE, Eq. (17). If one really needs the pressure itself from NSE we may resort to numeric integrations of the exact solution for the pressure. To compare this result, we take the x-derivative of the exact pressure solution from TE, and compare it with the NSE result from Eq. (17). Between the two results, we prefer the TE result for fundamental reasons. TE springs from fundamental physics, NSE uses a continuum engineering approach.

We thus suggest that the velocity field solutions from our time evolution equation, and the prescription of NSE for the pressure solves the 3-D Navier-Stokes Equation in a self-consistent way. The solutions are smooth, and there is no blowup time for the energy. But there is also no turbulence, as we have been proposing all along [3].

## 11. The role of viscosity in the time evolution equations and NSE

Viscosity is an adjustable parameter in NSE, but in our TE solutions, it does not show up. Why? The answer is that viscosity and pressure are built into TE solutions internally defined from kinetic theory, not from the continuum approach of NSE. They are not put into the equations of transport by hand. In fact in our 1997 paper, it is possible to "derive" NSE from TE by the gradual introduction of assumptions and approximations, sometimes self-contradictory. We "derived" NSE from TE with the following result in an obvious notation

$$\frac{\partial u_\alpha}{\partial t} + u_\alpha \frac{\partial u_\beta}{\partial r_\beta} + \frac{\partial \Pi_{\alpha\beta}}{\partial r_\beta} = v \frac{\partial^2 u_\alpha}{\partial r_\alpha \partial r_\alpha}.$$

(18)

with the following approximate definitions for the pressure

$$\Pi_{\beta\nu} = \frac{1}{m^2} \int_0^t ds\, s \int dp \int dr'\, f(r') \left( \frac{\partial V(r - ps/m - r')}{\partial r_\beta} \right) \left( \frac{p_\alpha p_\gamma}{m} \frac{\partial}{\partial r_\gamma} f(r - ps/m) \varphi(p) \right).$$

(19)

and the viscosity

$$v_{\alpha\beta} = \frac{1}{m}\int_0^t ds \int dr' \left(\frac{\partial V(r - ps/m - r')}{\partial r_\alpha}\right)\left(\frac{\partial V(r - ps/m - r')}{\partial r_\beta}\right) f(r') \frac{s^2}{m^2}$$

all defined for a non-uniform system with a factored one-particle distribution function $f(r,p,t) = n(r,t)\varphi(p,t)$. Note that if the system were homogeneous, at this stage of the approximations used in [1], the pressure will be zero, resulting in Burgers' equation, with no turbulent phenomenon.

### 12. Conclusions

Difficult non-linear partial differential equations are sometimes solved by inspired initial approximations, the origin of which may not always be clear. In our work, the initial approximation used is exact, the solution given by TE. The self-consistent calculation of the pressure simply follows. As a result, the 3-D Navier-Stokes may be considered solved exactly. But there is a price. We must give up the continuum model and adopt concepts from kinetic theory --molecules -- that which have been ignored all along. The coupling strength of the pair-potential $g$ enters the picture. Because of the existence of modern tools in the form of symbolic computation, as illustrated in some detail in this Report for the purpose of independent verification, other central pair potentials, or our function , $h(x,y,z)$ from Eq. (6) may be introduced. It will be possible to automate such procedure as we have outlined using symbolic computation. There will be as many solutions to NSE as there are computable pair-potentials, figuratively from no known solution to a feast of solutions. We also expect to see engineering applications of TE, especially when we relax some of the initial assumptions. We predict that the qualitative result will be the same, there will be no turbulence by any generous classical definition of the phenomenon. Observe that not even the Clay stipulations offer a definition of turbulence. Have we been searching for the phenomenon of turbulence without a precise mathematical definition? Contrast this with an approach described in [3], where it is suggested that the origin of turbulence lies in quantum mechanics.

**Appendix**

For the purpose of independent verifications, we complement the narrative in our Report with fragments of Maple codes which may be edited or combined to generate more insight.

The following Maple 13 code generates velocity field plots:

```
> gc(); restart;

> a := 'a'; g := 'g'; L := 'L';

>

> V := proc (x2, x) g*exp(-a*((x2-x)^2+(y2-y)^2+(z2-z)^2)) end proc;

>

> DVdx2 := diff(V(x2, x), x2);
```

```
> DVsq := DVdx2^2;

> IntDVsq := Int(DVsq, x2 = 0 .. L);

> IntDVsqx := int(DVsq, x2 = 0 .. L);

> IntDvsqx := simplify(IntDVsqx);

> Int2DVsqxy := int(IntDVsqx, y2 = 0 .. L);

> Int2DVsqxy := simplify(Int2DVsqxy);

> Int3DVsq := int(Int2DVsqxy, z2 = 0 .. L);

>

> Int3DVsqshifted := subs(x = x-px*s2/m, y = y-py*s2/m, z = z-pz*s2/m, Int3DVsq);

>

> Int3DVsq := 0;

> Int3DVsqshifted;

> gc(); Int3DVsqshifted;

>

> DVdx2 := 0;

> DVsq := 0;

> IntDVsq := 0;

> IntDVsqx := 0;

> IntDvsqx := 0;

> Int2DVsqxy := 0;

> Int2DVsqxy := 0;

> Int3DVsq := 0;

> gc();

> NULL;

> zmomentumpropagator := pz*exp(-px*s2*Dx/m-py*s2*Dy/m-pz*s2*Dz/m);
```

```
> Dpz1momentumpropagator := diff(zmomentumpropagator, pz);

> Dpz2momentumpropagator := diff(Dpz1momentumpropagator, pz);

> zmomentumchange := Dpz2momentumpropagator*'Int3DVsq';

> zmomentumchange := (-2*s2*Dz/m+pz*s2^2*Dz^2/m^2)*'Int3DVsqshifted';

> Int3DVsqshifted;

>

> zmomentumchange := -2*s2*(diff(Int3DVsqshifted, z))/m+pz*s2^2*(diff(diff(Int3DVsqshifted, z), z))/m^2;

>

> zmomentumchange := simplify(zmomentumchange);

> zmomentumchange := subs(py = 0, pz = 0, px = po, zmomentumchange);

>

> zmomentumchanges1 := int(zmomentumchange, s2 = 0 .. s1);

> zmomentumchanges1 := simplify(zmomentumchanges1);

> zmomentumchanget := int(zmomentumchanges1, s1 = 0 .. t);

> zmomentumchanget := simplify(zmomentumchanget);

> zmom := zmomentumchanget;

> zmomentumchanget := 0;

> zmomentumchanges1 := 0;

>

>

> zmom;

> NULL;

> NULL;
```

> xmomentumpropagator := px*exp(-px*s2*Dx/m-py*s2*Dy/m-pz*s2*Dz/m);

> Dpx1momentumpropagator := diff(xmomentumpropagator, px);

> Dpx2momentumpropagator := diff(Dpx1momentumpropagator, px);

> xmomentumchange := Dpx2momentumpropagator*'Int3DVsq';

> xmomentumchange := (-2*s2*Dx/m+pz*s2^2*Dx^2/m^2)*'Int3DVsqshifted';

> Int3DVsqshifted;

>

> xmomentumchange := -2*s2*(diff(Int3DVsqshifted, x))/m+px*s2^2*(diff(diff(Int3DVsqshifted, x), x))/m^2;

>

> xmomentumchange := simplify(xmomentumchange);

> xmomentumchange := subs(py = 0, pz = 0, px = po, xmomentumchange);

>

> xmomentumchanges1 := int(xmomentumchange, s2 = 0 .. s1);

> xmomentumchanges1 := simplify(xmomentumchanges1);

> xmomentumchanget := int(xmomentumchanges1, s1 = 0 .. t);

> xmomentumchanget := simplify(xmomentumchanget);

> xmomentumchanget;

> xmom := po+xmomentumchanget;

> xmom;

> xmomentumchanget := 0;

> xmomentumchange := simplify(xmomentumchange);

>

> xmom;

>

> xmomentumchanges1 := 0;

```
> xmomentumchanges1 := 0;

> xmomentumchanget := 0;

> xmomentumchanget := 0;

> xmomentumchanget := 0;

> xmom;

> zmom;

> ymoma := subs(z = uu, zmom);

> ymomb := y = z, ymoma;

> ymom := subs(uu = y, ymomb);

> ymoma := 0;

> ymomb := 0;

> gc();

> xmom;

> ymom;

> zmom;

>

>

> a := 'a'; g := 'g'; m := 'm'; po := 'po'; L := 'L';

> a := 1; g := 1; m := 1; po := 10; L := 10;

> with(plots);

> y := 'y'; z := 'z'; y := 0; z := 0;

> xmom;

> ymom;

> zmom;

> t := .1;
```

```
> xmom;

> ymom;

> zmom;

> t := .1; plot(xmom, x = 0 .. 10, axes = boxed, title = "Cartesian plot of x-momentum, y=z=0  t = 0.1"); t := 't';

> t := 2.0; plot(xmom, x = 0 .. 10, axes = boxed, title = "Cartesian plot of x-momentum , y=z=0 t = 2.0"); t := 't';

> y := 'y'; x := 'x'; t := 't'; t := .1; plot3d(xmom, x = 0 .. L, y = 0 .. L, axes = boxed, title = "x-momentum (x,y) t = 0.1");

> y := 'y'; x := 'x'; t := 't'; t := 2.0; plot3d(xmom, x = 0 .. L, y = 0 .. L, axes = boxed, title = "x-momentum (x,y) t = 2.0");

> t := .1; plot(ymom, x = 0 .. 10, axes = boxed, title = "y-momentum  t = 0.1"); t := 't';

> t := 2.0; plot(ymom, x = 0 .. 10, axes = boxed, title = "y-momentum  t = 2.0"); t := 't';

> y := 'y'; x := 'x'; t := 't'; t := .1; plot3d(ymom, x = 0 .. L, y = 0 .. L, axes = boxed, title = "y-momentum (x,y) t  = 0.1");

> y := 'y'; x := 'x'; t := 't'; t := 2.0; plot3d(xmom, x = 0 .. L, y = 0 .. L, axes = boxed, title = "y=momentum (x,y) t  = 2.0");

> t := 't'; y := 'y'; x := 'x'; z := 'z'; y := 0; z := 0; plot3d(xmom, x = 0 .. L, t = 0 .. 2, axes = boxed, title = "Time evolution of y-momentum");

> t := 't'; y := 'y'; x := 'x'; z := 'z'; y := 0; z := 0; plot3d(ymom, x = 0 .. L, t = 0 .. .2, axes = boxed, title = "Time evolution of x-momentum");

> x := 'x'; y := 'y'; z := 'z';

> xmom;

> t := 't'; theta := 'theta'; phi := 'phi'; omega := 'omega'; xmomperiodic := subs(x = L*sin(theta), y = L*sin(phi), z = L*sin(omega), xmom);

> omega := 0; t := 't'; t := 1.; plot3d(xmomperiodic, theta = 0 .. Pi, phi = 0 .. Pi, axes = boxed, title = "x-momentum in 3d toroid. t = 0.1 ");

> t := 't'; theta := 'theta'; phi := 'phi'; omega := 'omega'; ymomperiodic := subs(x = L*sin(theta), y = L*sin(phi), z = L*sin(omega), xmom);
```

```
> omega := 0; t := 't'; t := 200; plot3d(ymomperiodic, theta = 0 .. Pi, phi = 0 .. Pi, axes = boxed, title = "y-momentum in 3d toroid. t = 200");

> t := 't'; theta := 'theta'; phi := 'phi'; omega := 'omega'; zmomperiodic := subs(x = L*sin(theta), y = L*sin(phi), z = L*sin(omega), xmom);

> omega := 0; t := 't'; t := .1; plot3d(zmomperiodic, theta = 0 .. Pi, phi = 0 .. Pi, axes = boxed, title = "z-momentum in 3d toroid. t = 0.1 ");

> omega := 0; t := 't'; t := 20; plot3d(zmomperiodic, theta = 0 .. Pi, phi = 0 .. Pi, axes = boxed, title = "z-momentum in 3d toroid. t = 20");

>
```

The following Maple 13 code generates energy plots:

NULL

```
> gc();

> restart;

>

> a := 'a'; g := 'g'; L := 'L'; omega := 'omega'; theta := 'theta'; phi := 'phi'; po := 'po'; t := 't'; g := 'g'; a := 'a'; m := 'm';

>

> V := proc (x2, x) g*exp(-a*((x2-x)^2+(y2-y)^2+(z2-z)^2)) end proc;

>

> DVdx2 := diff(V(x2, x), x2);

> DVsq := DVdx2^2;

> IntDVsq := Int(DVsq, x2 = 0 .. L);

> IntDVsqx := int(DVsq, x2 = 0 .. L);

> IntDvsqx := simplify(IntDVsqx);
```

> Int2DVsqxy := int(IntDVsqx, y2 = 0 .. L);

> Int2DVsqxy := simplify(Int2DVsqxy);

> Int3DVsq := int(Int2DVsqxy, z2 = 0 .. L);

> Int3DVsq := simplify(Int3DVsq);

> L := 'L'; g := 'g'; a := 'a'; L := 'L'; g := 'g'; a := 'a'; po := 'po';

>

> Int3DVsqshifted := subs(x = x-px*s2/m, y = y-py*s2/m, z = z-pz*s2/m, Int3DVsq);

>

> xenergypropagator := px^2*exp(-px*s2*Dx/m-py*s2*Dy/m-pz*s2*Dz/m);

> Dpx1energypropagator := diff(xenergypropagator, px);

> Dpx2energypropagator := diff(Dpx1energypropagator, px);

> xenergychange := Dpx2energypropagator*'Int3DVsq';

> xenergychange := (2*exp(-px*s2*Dx/m-py*s2*Dy/m-pz*s2*Dz/m)-4*px*s2*Dx*exp(-px*s2*Dx/m-py*s2*Dy/m-pz*s2*Dz/m)/m+px^2*s2^2*Dx^2*exp(-px*s2*Dx/m-py*s2*Dy/m-pz*s2*Dz/m)/m^2)*'Int3DVsqshifted';

> Int3DVsqshifted;

> xenergychange := (2-4*px*s2*Dx/m+px^2*s2^2*Dx^2/m^2)*'Int3DVsqshifted';

> xenergychange := 2*Int3DVsqshifted-4*px*s2*(diff(Int3DVsqshifted, x))/m+px^2*s2^2*(diff(diff(Int3DVsqshifted, x), x))/m^2;

> xenergychange := simplify(xenergychange);

> xenergychange := subs(py = 0, pz = 0, px = po, xenergychange);

>

> xenergychanges1 := int(xenergychange, s2 = 0 .. s1);

> xenergychanges1 := simplify(xenergychanges1);

> xenergychanget := int(xenergychanges1, s1 = 0 .. t);

>

> xenergychanget := simplify(xenergychanget);

```
> xenergy := po^2/(2*m)+(1/2)*xenergychanget;

> zenergypropagator := pz^2*exp(-px*s2*Dx/m-py*s2*Dy/m-pz*s2*Dz/m);

> Dpzenergypropagator := diff(zenergypropagator, pz);

> Dpz2zenergypropagator := diff(Dpzenergypropagator, pz);

> zenergyychange := Dpz2zenergypropagator*'Int3DVsq';

> zenergychange := (2-4*pz*s2*Dy/m+pz^2*s2^2*Dy^2/m^2)*'Int3DVsqshifted';

> Int3DVsqshifted;

>

> zenergychange := simplify(zenergychange);

> zenergychange := subs(py = 0, pz = 0, px = po, zenergychange);

>

> zenergychanges1 := int(zenergychange, s2 = 0 .. s1);

> zenergychanges1 := simplify(zenergychanges1);

> zenergychanget := int(zenergychanges1, s1 = 0 .. t);

> zenergychanget := simplify(zenergychanget);

> zenergy := (1/2)*zenergychanget;

> yenergya := subs(z = uu, zenergy);

> yenergyb := subs(y = z, yenergya);

> yenergy := subs(uu = y, yenergyb);

> yenergy;

>

> zenergy;

> xenergy;

> yenergy;

> zenergy;
```

```
> 
> a := 'a'; g := 'g'; po := 'po'; L := 'L'; m := 'm'; a := 1; g := 1; po := 10; L := 10; m := 1;

> xenergy;

> yenergy;

> zenergy;

> with(plots); xenergy;

> t := 't'; t := .1; z := 0; plot3d(xenergy, x = 0 .. L, y = 0 .. L, axes = boxed, title = "x-energy as a function of x, y   t =0.1"); t := 't'; z := 'z';

> t := 't'; t := .2; z := 0; plot3d(xenergy, x = 0 .. L, y = 0 .. L, axes = boxed, title = "x-energy as a function of x, y   t =  0.2"); t := 't'; z := 'z';

> t := 't'; t := .1; z := 0; plot3d(yenergy, x = 0 .. L, y = 0 .. L, axes = boxed, title = "y-energy as a function of x, y   t =0.1"); t := 't'; z := 'z';

> t := 't'; t := .2; z := 0; plot3d(xenergy, x = 0 .. L, y = 0 .. L, axes = boxed, title = "x-energy as a function of x, y   t =0.2"); t := 't'; z := 'z';

> t := 't'; t := .1; z := 0; plot3d(yenergy, x = 0 .. L, y = 0 .. L, axes = boxed, title = "y-energy as a function of x, y   t =0.1"); t := 't'; z := 'z';

> t := 't'; t := .1; z := 0; plot3d(zenergy, x = 0 .. L, y = 0 .. L, axes = boxed, title = "z-energy as a function of x, y   t =0.1"); t := 't'; z := 'z';

> NULL;

> t := 't'; t := .1; z := 0; plot3d(4*xenergy, x = 0 .. L, y = 0 .. L, axes = boxed, title = "x-pressure as a function of  x, y   t =0.1"); t := 't'; z := 'z';

> t := 't'; t := 't'; z := 0; y := 0; xpressure := 4*xenergy; plot3d(4*xenergy, x = 0 .. L, t = 0 .. 2, axes = boxed, title = "x-pressure as a function of  t "); t := 't'; z := 'z';

> t := 't'; t := .1; z := 0; plot3d(2*yenergy, x = 0 .. L, y = 0 .. L, axes = boxed, title = "y-pressure as a function of  x, y   t =0.1"); t := 't'; z := 'z';

> t := 't'; t := .2; z := 0; plot3d(2*xenergy, x = 0 .. L, y = 0 .. L, axes = boxed, title = "x-pressure as a function of  x, y   t =0.2"); t := 't'; z := 'z';

> t := 't'; t := .1; z := 0; plot3d(2*yenergy, x = 0 .. L, y = 0 .. L, axes = boxed, title = "y-pressure as a function of  x, y   t =0.1"); t := 't'; z := 'z';
```

```
> t := 't'; t := .1; z := 0; plot3d(2*zenergy, x = 0 .. L, y = 0 .. L, axes = boxed, title = "z-pressure as a function of x, y  t =0.1"); t := 't'; z := 'z';

> a := 'a'; g := 'g'; m := 'm'; L := 'L'; po := 'po'; nu := 'nu';

> t := 't'; xpressure := 4*xenergy;

> xpressure;

> a := 1; g := 1; m := 1; L := 10; nu := 1; po := 10;

> y := 0; z := 0; x := 'x'; t := 't'; xpressure; plot3d(xpressure, x = 0 .. L, t = 0 .. 2, axes = boxed, title = "x-pressure from TE");

> DPdxTE := diff(xpressure, x);

> y := 0; z := 0; x := 'x'; t := 't'; xpressure; plot3d(DPdxTE, x = 0 .. L, t = 0 .. 2, axes = boxed, title = "gradient of x-pressure from TE");

> t := 't'; t := .1; z := 0; plot3d(xenergy+yenergy+zenergy, x = 0 .. L, y = 0 .. L, axes = boxed, title = "total energy as a function of x, y  t =0.1"); t := 't'; z := 'z';

> t := 't'; t := .2; z := 0; plot3d(xenergy+yenergy+zenergy, x = 0 .. L, y = 0 .. L, axes = boxed, title = "total energy as a function of x, y  t =0.2"); t := 't'; z := 'z';

> t := 't'; x := 'x'; y := 'y'; z := 'z';

> a := 'a'; g := 'g'; m := 'm'; L := 'L';

> xenergy;

> yenergy;

> zenergy;

> x := 'x'; y := 'y'; z := 'z';

> xenergytoroid := subs(x = L*sin(theta), y = L*sin(phi), z = L*sin(omega), xenergy);

> yenergytoroid := subs(x = L*sin(theta), y = L*sin(phi), z = L*sin(omega), yenergy);

> zenergytoroid := subs(x = L*sin(theta), y = L*sin(phi), z = L*sin(omega), zenergy);

> omega := 'omega'; omega := 0; theta := 'theta'; phi := 'phi';

> with(plots);

> m := 'm'; g := 'g'; L := 'L'; po := 'po'; a := 'a'; m := 1; g := 1; L := 10; po := 10; a := 1;
```

> 

> t := 't'; t := .2; omega := 'omega'; omega := 0; plot3d(xenergytoroid, theta = 0 .. Pi, phi = 0 .. Pi, axes = boxed, title = " x-energy as function of theta, phi,t =0. 2");

> xenergytoroid;

> t := 't'; t := .2; omega := 'omega'; omega := 0; plot3d(xenergytoroid, theta = 0 .. Pi, phi = 0 .. Pi, axes = boxed, title = " x-energy as function of theta, phi,t = 0.2");

> 

> totalenergytoroid;

> t := 't'; t := .2; omega := 'omega'; omega := 0; plot3d(xenergytoroid+yenergytoroid+zenergytoroid, theta = 0 .. Pi, phi = 0 .. Pi, axes = boxed, title = " total energy as function of theta, phi,t = 0.2");

> xenergypropagator := px^2*exp(-px*s2*Dx/m-py*s2*Dy/m-pz*s2*Dz/m);

> Dpx1energypropagator := diff(xenergypropagator, px);

> Dpx2energypropagator := diff(Dpx1energypropagator, px);

> xenergychange := Dpx2energypropagator*'Int3DVsq';

> xenergychange := (2*exp(-px*s2*Dx/m-py*s2*Dy/m-pz*s2*Dz/m)-4*px*s2*Dx*exp(-px*s2*Dx/m-py*s2*Dy/m-pz*s2*Dz/m)/m+px^2*s2^2*Dx^2*exp(-px*s2*Dx/m-py*s2*Dy/m-pz*s2*Dz/m)/m^2)*'Int3DVsqshifted';

> Int3DVsqshifted;

> xenergychange := (2-4*px*s2*Dx/m+px^2*s2^2*Dx^2/m^2)*'Int3DVsqshifted';

> xenergychange := 2*Int3DVsqshifted-4*px*s2*(diff(Int3DVsqshifted, x))/m+px^2*s2^2*(diff(diff(Int3DVsqshifted, x), x))/m^2;

> xenergychange := simplify(xenergychange);

> xenergychange := subs(py = 0, pz = 0, px = po, xenergychange);

> 

> xenergychanges1 := int(xenergychange, s2 = 0 .. s1);

> xenergychanges1 := simplify(xenergychanges1);

> xenergychanget := int(xenergychanges1, s1 = 0 .. t);

```
> 
> xenergychanget := simplify(xenergychanget);
> xenergy := po^2/(2*m)+(1/2)*xenergychanget;
> xpressure := 4*xenergy;
> DPxTE := diff(xpressure, x);
> 
> a := 'a'; g := 'g'; m := 'm'; L := 'L'; po := 'po'; nu := 'nu'; t := 't'; x := 'x'; y := 'y'; z := 'z';
> a := 1; g := 1; m := 1; L := 10; po := 10; nu := 1; y := 0; z := 0;
> DPxTE;
> with(plots); plot3d(DPxTE, x = 0 .. L, t = 0 .. 2.0, axes = boxed, title = "x-gradienr of pressure from TE");
> a := 'a'; g := 'g'; m := 'm'; L := 'L'; po := 'po'; nu := 'nu'; t := 't'; x := 'x'; y := 'y'; z := 'z';
> zenergypropagator := pz^2*exp(-px*s2*Dx/m-py*s2*Dy/m-pz*s2*Dz/m);
> Dpzenergypropagator := diff(zenergypropagator, pz);
> Dpz2zenergypropagator := diff(Dpzenergypropagator, pz);
> zenergyychange := Dpz2zenergypropagator*'Int3DVsq';
> zenergychange := (2-4*pz*s2*Dy/m+pz^2*s2^2*Dy^2/m^2)*'Int3DVsqshifted';
> Int3DVsqshifted;
> 
> zenergychange := simplify(zenergychange);
> zenergychange := subs(py = 0, pz = 0, px = po, zenergychange);
> 
> zenergychanges1 := int(zenergychange, s2 = 0 .. s1);
> zenergychanges1 := simplify(zenergychanges1);
> zenergychanget := int(zenergychanges1, s1 = 0 .. t);
> zenergychanget := simplify(zenergychanget);
```

```
> zenergy := (1/2)*zenergychanget;

> yenergya := subs(z = uu, zenergy);

> yenergyb := subs(y = z, yenergya);

> yenergy := subs(uu = y, yenergyb);

>

> xenergy;

> yenergy;

> zenergy;

>

> a := 'a'; g := 'g'; po := 'po'; L := 'L'; m := 'm'; a := 1; g := 1; po := 10; L := 10; m := 1;

> xenergy;

> yenergy;

> zenergy;

> with(plots); xenergy;

> t := 't'; t := .1; z := 0; plot3d(xenergy, x = 0 .. L, y = 0 .. L, axes = boxed, title = "x-energy as
       a function of  x, y   t =0.1"); t := 't'; z := 'z';

> t := 't'; t := .2; z := 0; plot3d(xenergy, x = 0 .. L, y = 0 .. L, axes = boxed, title = "x-energy as
       a function of  x, y   t =  0.2"); t := 't'; z := 'z';

> t := 't'; t := .1; z := 0; plot3d(yenergy, x = 0 .. L, y = 0 .. L, axes = boxed, title = "y-energy as
       a function of  x, y   t =0.1"); t := 't'; z := 'z';

> t := 't'; t := .2; z := 0; plot3d(xenergy, x = 0 .. L, y = 0 .. L, axes = boxed, title = "x-energy as
       a function of  x, y   t =0.2"); t := 't'; z := 'z';

> t := 't'; t := .1; z := 0; plot3d(yenergy, x = 0 .. L, y = 0 .. L, axes = boxed, title = "y-energy as
       a function of  x, y   t =0.1"); t := 't'; z := 'z';

> t := 't'; t := .1; z := 0; plot3d(zenergy, x = 0 .. L, y = 0 .. L, axes = boxed, title = "z-energy as
       a function of  x, y   t =0.1"); t := 't'; z := 'z';

> NULL;
```

```
> t := 't'; t := .1; z := 0; plot3d(4*xenergy, x = 0 .. L, y = 0 .. L, axes = boxed, title = "x-
         pressure as a function of x, y  t =0.1"); t := 't'; z := 'z';

> t := 't'; t := 't'; z := 0; y := 0; xpressure := 4*xenergy; plot3d(4*xenergy, x = 0 .. L, t = 0 .. 2,
         axes = boxed, title = "x-pressure as a function of t "); t := 't'; z := 'z';

> t := 't'; t := .1; z := 0; plot3d(2*yenergy, x = 0 .. L, y = 0 .. L, axes = boxed, title = "y-
         pressure as a function of x, y  t =0.1"); t := 't'; z := 'z';

> t := 't'; t := .2; z := 0; plot3d(2*xenergy, x = 0 .. L, y = 0 .. L, axes = boxed, title = "x-
         pressure as a function of x, y  t =0.2"); t := 't'; z := 'z';

> t := 't'; t := .1; z := 0; plot3d(2*yenergy, x = 0 .. L, y = 0 .. L, axes = boxed, title = "y-
         pressure as a function of x, y  t =0.1"); t := 't'; z := 'z';

> t := 't'; t := .1; z := 0; plot3d(2*zenergy, x = 0 .. L, y = 0 .. L, axes = boxed, title = "z-
         pressure  as a function of x, y  t =0.1"); t := 't'; z := 'z';

> a := 'a'; g := 'g'; m := 'm'; L := 'L'; po := 'po'; nu := 'nu';

> t := 't'; xpressure := 4*xenergy;

> xpressure;

> a := 1; g := 1; m := 1; L := 10; nu := 1; po := 10;

> y := 0; z := 0; x := 'x'; t := 't'; xpressure; plot3d(xpressure, x = 0 .. L, t = 0 .. 2, axes = boxed,
         title = "x-pressure from TE");

> DPdxTE := diff(xpressure, x);

> y := 0; z := 0; x := 'x'; t := 't'; xpressure; plot3d(DPdxTE, x = 0 .. L, t = 0 .. 2, axes = boxed,
         title = "gradient of x-pressure from TE");

> t := 't'; t := .1; z := 0; plot3d(xenergy+yenergy+zenergy, x = 0 .. L, y = 0 .. L, axes = boxed,
         title = "total energy as a function of x, y  t =0.1"); t := 't'; z := 'z';

> t := 't'; t := .2; z := 0; plot3d(xenergy+yenergy+zenergy, x = 0 .. L, y = 0 .. L, axes = boxed,
         title = "total energy as a function of x, y  t =0.2"); t := 't'; z := 'z';

> t := 't'; x := 'x'; y := 'y'; z := 'z';

> a := 'a'; g := 'g'; m := 'm'; L := 'L';

> xenergy;

> yenergy;
```

```
> zenergy;

> xenergytoroid := subs(x = L*sin(theta), y = L*sin(phi), z = L*sin(omega), xenergy);

> yenergytoroid := subs(x = L*sin(theta), y = L*sin(phi), z = L*sin(omega), yenergy);

> zenergytoroid := subs(x = L*sin(theta), y = L*sin(phi), z = L*sin(omega), zenergy);

> omega := 'omega'; omega := 0; theta := 'theta'; phi := 'phi';

> with(plots);

> m := 'm'; g := 'g'; L := 'L'; po := 'po'; a := 'a'; m := 1; g := 1; L := 10; po := 10; a := 1;

>

>

> t := 't'; t := .1; omega := 'omega'; omega := 0; plot3d(xenergytoroid, theta = 0 .. Pi, phi =
        0 .. Pi, axes = boxed, title = " x-energy as function of theta, phi,t = 0.1");

> xenergytoroid;

> t := 't'; t := .2; omega := 'omega'; omega := 0; plot3d(xenergytoroid, theta = 0 .. Pi, phi =
        0 .. Pi, axes = boxed, title = " x-energy as function of theta, phi,t = 0.2");

>

>

> totalenergytoroid;

> t := 't'; t := .2; omega := 'omega'; omega := 0;
        plot3d(xenergytoroid+yenergytoroid+zenergytoroid, theta = 0 .. Pi, phi = 0 .. Pi,
        axes = boxed, title = " total energy as function of theta, phi,t = 0.2");

> t := 't'; t := 2.0; omega := 'omega'; omega := 0;
        plot3d(xenergytoroid+yenergytoroid+zenergytoroid, theta = 0 .. Pi, phi = 0 .. Pi,
        axes = boxed, title = " total energy as function of theta, phi, t = 2.0");

>
```

**The following Maple code generates the NSE pressure gradient in x**

```
>
>  gc( ); restart;
```

> $a := 'a'; g := 'g'; L := 'L'; m := 'm'; po := 'po'; \text{nu} := '\text{nu}'; x := 'x'; y := 'y'; z := 'z'; t := 't';$

>
> $V := \mathbf{proc}(x2, x) \quad g \cdot \exp\left(-a \cdot \left((x2 - x)^2 + (y2 - y)^2 + (z2 - z)^2\right)\right) \quad \mathbf{end\ proc};$

>
> $DVdx2 := \text{diff}(V(x2, x), x2);$
> $DVsq := (DVdx2)^2;$
> $IntDVsq := \text{Int}(DVsq, x2 = 0..L);$
> $IntDVsqx := \text{int}(DVsq, x2 = 0..L);$
> $IntDvsqx := \text{simplify}(IntDVsqx);$
> $Int2DVsqxy := \text{int}(IntDVsqx, y2 = 0..L);$
> $Int2DVsqxy := \text{simplify}(Int2DVsqxy);$
> $Int3DVsq := \text{int}(Int2DVsqxy, z2 = 0..L);$
>
> $Int3DVsqshifted := \text{subs}\left(x = x - \frac{px \cdot s2}{m}, y = y - \frac{py \cdot s2}{m}, z = z - \frac{pz \cdot s2}{m}, Int3DVsq\right);$

>
> $Int3DVsq := 0;$
> $Int3DVsqshifted;$
> $gc(); Int3DVsqshifted;$
>
>
>
> $DVdx2 := 0;$
> $DVsq := 0;$
> $IntDVsq := 0;$
> $IntDVsqx := 0;$
> $IntDvsqx := 0;$
> $Int2DVsqxy := 0;$
> $Int2DVsqxy := 0;$
> $Int3DVsq := 0;$
> $gc();$
> # momentum
> $zmomentumpropagator := pz \cdot \exp\left(-\frac{px \cdot s2}{m} Dx - \frac{py \cdot s2}{m} Dy - \frac{pz \cdot s2}{m} Dz\right);$

> $Dpz1momentumpropagator := \text{diff}(zmomentumpropagator, pz);$
> $Dpz2momentumpropagator := \text{diff}(Dpz1momentumpropagator, pz);$
> $zmomentumchange := Dpz2momentumpropagator \cdot 'Int3DVsq';$

> $zmomentumchange := \left(-\dfrac{2\,s2\,Dz}{m} + \dfrac{pz\,s2^2\,Dz^2}{m^2}\right)$ 'Int3DVsqshifted';

> Int3DVsqshifted;
>
> $zmomentumchange := \left(-\dfrac{2\,s2\,\mathit{diff}(\mathit{Int3DVsqshifted},z)}{m} + \dfrac{pz\,s2^2\,\mathit{diff}(\mathit{diff}(\mathit{Int3DVsqshifted},z),z)}{m^2}\right);$

>
> zmomentumchange := simplify(zmomentumchange);
> zmomentumchange := subs(py = 0, pz = 0, px = po, zmomentumchange);

>
>
>
> zmomentumchanges1 := int(zmomentumchange, s2 = 0 .. s1);
> zmomentumchanges1 := simplify(zmomentumchanges1);
> zmomentumchanget := int(zmomentumchanges1, s1 = 0 .. t);
> zmomentumchanget := simplify(zmomentumchanget);
> zmom := zmomentumchanget;

>
>
> zmomentumchanget := 0;
> zmomentumchanges1 := 0;
>
>
> zmom;
> ## x axis
> # momentum
> $xmomentumpropagator := px \cdot \exp\!\left(-\dfrac{px \cdot s2}{m}Dx - \dfrac{py \cdot s2}{m}Dy - \dfrac{pz \cdot s2}{m}Dz\right);$

> Dpx1momentumpropagator := diff(xmomentumpropagator, px);
> Dpx2momentumpropagator := diff(Dpx1momentumpropagator, px);

> xmomentumchange := Dpx2momentumpropagator · 'Int3DVsq';
> $xmomentumchange := \left(-\dfrac{2\,s2\,Dx}{m} + \dfrac{pz\,s2^2\,Dx^2}{m^2}\right)$ 'Int3DVsqshifted';

> Int3DVsqshifted;
>

> $xmomentumchange := \left( -\frac{2\, s2\, \text{diff}(Int3DVsqshifted, x)}{m} \right.$
  $\left. + \frac{px\, s2^2 \text{diff}(\text{diff}(Int3DVsqshifted, x), x\ )}{m^2} \right);$

>
> $xmomentumchange := simplify(xmomentumchange);$
> $xmomentumchange := subs(py = 0, pz = 0, px = po,$
  $xmomentumchange);$

>
>
>
>
> $xmomentumchanges1 := int(xmomentumchange, s2 = 0..s1);$
> $xmomentumchanges1 := simplify(xmomentumchanges1);$
> $xmomentumchanget := int(xmomentumchanges1, s1 = 0..t);$
> $xmomentumchanget := simplify(xmomentumchanget);$
> $xmomentumchanget;$
> $xmom := po + xmomentumchanget;$
> $xmom;$
> $xmomentumchanget := 0;$
> $xmomentumchange := simplify(xmomentumchange);$
>
> $xmom;$
>
>
>
> $xmomentumchanges1 := 0;$
> $xmomentumchanges1 := 0;$
> $xmomentumchanget := 0;$
> $xmomentumchanget := 0;$
> $xmomentumchanget := 0;$
> $xmom;$
> $zmom;$
> $ymoma := subs(z = uu, zmom);$
> $ymomb := (y = z, ymoma);$
> $ymom := subs(uu = y, ymomb);$
> $ymoma := 0;$
> $ymomb := 0;$
> $gc();$
> $xmom;$
> $ymom;$
> $zmom;$
> $Dxmomdt := diff(xmom, t);$
> $Dxmomdx := diff(xmom, x);$
> $Dxmomdx2 := diff(Dxmomdx, x);$
> $Dxmomdy := diff(xmom, y);$

> $Dxmomdy2 := \text{diff}(Dxmomdy, x);$
> $Dxmomdz := \text{diff}(xmom, z);$
> $Dxmomdz2 := \text{diff}(Dxmomdz, z);$
> $gradx := xmom \cdot Dxmomdx + ymom \cdot Dxmomdy + zmom \cdot Dxmomdz;$
> $Dxmomdx := 0; Dxmomdy := 0; Dxmomdz := 0; gc();$
>
> $xnabla := \nu \cdot Dxmomdx2 + \nu \cdot Dxmomdy2 + \nu \cdot Dxmomdz2;$
>
> $Dxmomdx2 := 0;$
>
> $Dxmomdy2 := 0;$
>
> $Dxmomdz2 := 0;$
>
>
> $DPx := xnabla - gradx - Dxmomdt;$
> $gradx := 0; xnabla := 0; Dxmomdt := 0;$
>
> $gc();$
> $DPx;$
> $a := 'a'; g := 'g'; m := 'm'; L := 'L'; po := 'po'; nu := 'nu'; a := 1; g := 1; m := 1; L := 10; po := 10; nu := 1;$
>
> $y := 'y'; z := 'z'; x := 'x'; y := 0; z := 0; t := 't';$
> $DPx;$
> $\text{with}(plots); \text{plot3d}(DPx, x = 0 .. L, t = 0 .. 2, axes = boxed, title = \text{"x-gradient of pressure from NSE"});$
>

```
> gc(); restart;
> a := 'a'; g := 'g'; L := 'L';
>
> V := proc (x2, x) g*exp(-a*((x2-x)^2+(y2-y)^2+(z2-z)^2)) end proc;
>
> DVdx2 := diff(V(x2, x), x2);
> DVsq := DVdx2^2;
> IntDVsq := Int(DVsq, x2 = 0 .. L);
> IntDVsqx := int(DVsq, x2 = 0 .. L);
> IntDvsqx := simplify(IntDVsqx);
> Int2DVsqxy := int(IntDVsqx, y2 = 0 .. L);
> Int2DVsqxy := simplify(Int2DVsqxy);
> Int3DVsq := int(Int2DVsqxy, z2 = 0 .. L);
>
> Int3DVsqshifted := subs(x = x-px*s2/m, y = y-py*s2/m, z = z-pz*s2/m, Int3DVsq);
>
> Int3DVsq := 0;
> Int3DVsqshifted;
> gc(); Int3DVsqshifted;
>
>
> DVdx2 := 0;
> DVsq := 0;
> IntDVsq := 0;
```

```
> IntDVsqx := 0;

> IntDvsqx := 0;

> Int2DVsqxy := 0;

> Int2DVsqxy := 0;

> Int3DVsq := 0;

> gc();

> NULL;

>

> xmomentumpropagator := px*exp(-px*s2*Dx/m-py*s2*Dy/m-pz*s2*Dz/m);

> Dpx1momentumpropagator := diff(xmomentumpropagator, px);

> Dpx2momentumpropagator := diff(Dpx1momentumpropagator, px);

> xmomentumchange := Dpx2momentumpropagator*'Int3DVsq';

> xmomentumchange := (-2*s2*Dx/m+pz*s2^2*Dx^2/m^2)*'Int3DVsqshifted';

> Int3DVsqshifted;

>

> xmomentumchange := -2*s2*(diff(Int3DVsqshifted, x))/m+px*s2^2*(diff(diff(Int3DVsqshifted, x), x))/m^2;

>

> xmomentumchange := simplify(xmomentumchange);

> xmomentumchange := subs(py = 0, pz = 0, px = po, xmomentumchange);

>

> xmomentumchanges1 := int(xmomentumchange, s2 = 0 .. s1);

> xmomentumchanges1 := simplify(xmomentumchanges1);

> xmomentumchanget := int(xmomentumchanges1, s1 = 0 .. t);

> xmomentumchanget := simplify(xmomentumchanget);

> xmomentumchanget;
```

```
> xmom := po+xmomentumchanget;

> xmom;

> xmomentumchanget := 0;

>

> xmom;

>

> xmomentumchanges1 := 0;

> xmomentumchanges1 := 0;

> xmomentumchanget := 0;

> xmomentumchanget := 0;

> xmomentumchanget := 0;

> xmom;

> NULL;

> zmomentumpropagator := pz*exp(-px*s2*Dx/m-py*s2*Dy/m-pz*s2*Dz/m);

> Dpz1momentumpropagator := diff(zmomentumpropagator, pz);

> Dpz2momentumpropagator := diff(Dpz1momentumpropagator, pz);

> zmomentumchange := Dpz2momentumpropagator*'Int3DVsq';

> zmomentumchange := (-2*s2*Dz/m+pz*s2^2*Dz^2/m^2)*'Int3DVsqshifted';

> Int3DVsqshifted;

>

> zmomentumchange := -2*s2*(diff(Int3DVsqshifted, z))/m+pz*s2^2*(diff(diff(Int3DVsqshifted, z), z))/m^2;

>

> zmomentumchange := simplify(zmomentumchange);

> zmomentumchange := subs(py = 0, pz = 0, px = po, zmomentumchange);

>
```

```
> 
> 
> zmomentumchanges1 := int(zmomentumchange, s2 = 0 .. s1);

> zmomentumchanges1 := simplify(zmomentumchanges1);

> zmomentumchanget := int(zmomentumchanges1, s1 = 0 .. t);

> zmomentumchanget := simplify(zmomentumchanget);

> zmom := zmomentumchanget;

> 
> zmomentumchanget := 0;

> zmomentumchanges1 := 0;

> 
> 
> zmom;

> 
> 
> NULL;

> ymomentumpropagator := py*exp(-px*s2*Dx/m-py*s2*Dy/m-pz*s2*Dz/m);

> Dpy1momentumpropagator := diff(ymomentumpropagator, py);

> Dpy2momentumpropagator := diff(Dpy1momentumpropagator, py);

> ymomentumchange := Dpy2momentumpropagator*'Int3DVsq';

> ymomentumchange := (-2*s2*Dz/m+py*s2^2*Dz^2/m^2)*'Int3DVsqshifted';

> Int3DVsqshifted;

> 
> ymomentumchange := -2*s2*(diff(Int3DVsqshifted, z))/m+py*s2^2*(diff(diff(Int3DVsqshifted, z), z))/m^2;
```

```
> 
> ymomentumchange := simplify(ymomentumchange);
> ymomentumchange := subs(py = 0, pz = 0, px = po, ymomentumchange);
> 
> ymomentumchanges1 := int(ymomentumchange, s2 = 0 .. s1);
> ymomentumchanges1 := simplify(ymomentumchanges1);
> ymomentumchanget := int(ymomentumchanges1, s1 = 0 .. t);
> ymomentumchanget := simplify(ymomentumchanget);
> ymom := ymomentumchanget;

> 
> ymomentumchanget := 0;
> ymomentumchanges1 := 0;
> x := 'x'; y := 'y'; z := 'z'; t := 't';
> a := 'a'; g := 'g'; m := 'm'; L := 'L'; po := 'po';
> xmom;
> zmom;
> xmom;
> NULL;
> with(plots);
> a := 'a'; g := 'g'; m := 'm'; L := 'L'; po := 'po';
> 
> a := 1; g := 1; m := 1; L := 10; po := 10;
> xmom; xmom := simplify(xmom);
> y := 0; z := 0; T := 'T'; T := 2; plot3d(xmom, x = 0 .. 10, t = 0 .. T, axes = boxed, title = "x-momentum as a function of t"); T := 'T';
```

> zmom; zmom := simplify(zmom);

> y := 0; z := 0; T := 'T'; T := .2; plot3d(zmom, x = 0 .. 10, t = 0 .. T, axes = boxed, title = "z-momentum as a function of t"); T := 'T';

> y := 0; z := 0; T := 'T'; T := 1.0; plot3d(xmom, x = 0 .. 10, t = 0 .. T, axes = boxed, title = "x-momentum as a function of t"); T := 'T';

> y := 0; z := 0; T := 'T'; T := 2; plot3d(xmom, x = 0 .. 10, t = 0 .. T, axes = boxed, title = "x-momentum as a function of t"); T := 'T';

> zmom; zmom := simplify(zmom);

> y := 0; z := 0; T := 'T'; T := .3; plot3d(zmom, x = 0 .. 10, t = 0 .. T, axes = boxed, title = "z-momentum as a function of t"); T := 'T';